\documentclass{emulateapj}

\usepackage{color}
\definecolor{dr}{rgb}{0.6,0,0}
\definecolor{db}{rgb}{0,0,0.6}

\usepackage[ hyperfootnotes={false},
dvips, pdftitle={Manuscript}, pdfauthor={Yue Shen},
pdfstartview={FitH}, bookmarksopen={true} ]{hyperref}
\hypersetup{colorlinks={true}, urlcolor={db}, citecolor={dr}}

\usepackage{lscape}
\usepackage{apjfonts}
\usepackage{graphicx}
\usepackage{natbib}
\begin{document}

\title{Supermassive Black Holes in the Hierarchical Universe:
A General Framework and Observational Tests}

\shorttitle{SMBHS IN THE HIERARCHICAL UNIVERSE}

\slugcomment{Submitted to ApJ}

\shortauthors{SHEN}
\author{Yue Shen\\
Princeton University Observatory, Princeton, NJ 08544.}

\begin{abstract}
We present a simple framework for the growth and evolution of
supermassive black holes (SMBHs) in the hierarchical structure
formation paradigm, adopting the general idea that quasar activity
is triggered in major mergers. In our model, black hole accretion
is triggered during major mergers (mass ratio $\gtrsim 0.3$)
between host dark matter halos. The successive evolution of quasar
luminosities follows a universal light curve form, during which
the growth of the SMBH is modeled self-consistently: an initial
exponential growth at constant Eddington ratio of order unity
until it reaches the peak luminosity, followed by a power-law
decay. Assuming that the peak luminosity correlates with the
post-merger halo mass, we convolve the light curve with the
triggering rate of quasar activity to predict the quasar
luminosity function (LF). Our model reproduces the observed LF at
$0.5<z<4.5$ for the full luminosity ranges probed by current
optical and X-ray surveys. At $z<0.5$, our model underestimates
the LF at $L_{\rm bol}<10^{45}\,{\rm ergs^{-1}}$, allowing room
for AGN activity triggered by secular processes instead of major
mergers. At $z>4.5$, in order to reproduce the observed quasar
abundance, the typical quasar hosts must shift to lower mass
halos, and/or minor mergers can also trigger quasar activity. Our
model reproduces both the observed redshift evolution and
luminosity dependence of the linear bias of quasar/AGN clustering.
Due to the scatter between instantaneous luminosity and halo mass,
quasar/AGN clustering weakly depends on luminosity at low to
intermediate luminosities; but the linear bias rises rapidly with
luminosity at the high luminosity end and at high redshift. In our
model, the Eddington ratio distribution is roughly log-normal,
which broadens and shifts to lower mean values from high
luminosity quasars ($L_{\rm bol}\gtrsim 10^{46}\,{\rm ergs^{-1}}$)
to low luminosity AGNs ($L_{\rm bol}\lesssim 10^{45}\,{\rm
ergs^{-1}}$), in good agreement with observations. The model
predicts that the vast majority of $\gtrsim 10^{8.5}\,M_\odot$
SMBHs were already in place by $z=1$, and $\lesssim 50\%$ of them
were in place by $z=2$; but the less massive ($\lesssim 10^{7}\,
M_\odot$) SMBHs were assembled more recently, likely more through
secular processes than by major mergers -- in accordance with the
downsizing picture of SMBH assembly since the peak of bright
quasar activity around $z\sim 2-3$.

\end{abstract}

\keywords{black hole physics -- galaxies: active -- cosmology:
observations -- large-scale structure of universe -- quasars:
general -- surveys}

\section{Introduction}
The cosmic assembly and evolution of supermassive black holes
(SMBHs) is a central topic in modern cosmology: in the local
universe, SMBHs reside in almost every bulge dominant galaxy
\citep[e.g.,][]{Kormendy_Richstone_1995,Richstone_etal_1998}; and
they likely played important roles during their coevolution with
galaxy bulges
\citep[e.g.,][]{Silk_Rees_1998,Wyithe_Loeb_2002,Wyithe_Loeb_2003,
King_2003,Di_Matteo_etal_2005,Croton_etal_2006}, as inferred from
observed scaling relations between bulge properties and the mass
of the central BH
\citep[e.g.,][]{Magorrian_etal_1998,Ferrarese_Merritt_2000,
Gebhardt_etal_2000a, Graham_etal_2001,
Tremaine_etal_2002,Marconi_Hunt_2003}. It is also generally
accepted that the local dormant SMBH population was largely
assembled via gas accretion during a luminous QSO
phase\footnote{In this paper we use the term ``QSO'' to refer to
all actively accreting SMBHs. We adopt the convention that $L_{\rm
bol}=10^{45}\,{\rm ergs^{-1}}$ is the dividing line between
quasars and AGNs. This division is slightly lower than the
traditional Seyfert/quasar division \citep[][]{Schmidt_Green_1983}
of $M_{B}=-23$ or $L_{\rm bol}\sim 10^{12}\,L_\odot$.}
\citep[e.g.,][]{Salpeter_1964,Zeldovich_Novikov_1964,Lynden-Bell_1969,Soltan_1982,
Small_Blandford_1992,Salucci_etal_1999,Yu_Tremaine_2002,Shankar_etal_2004,Marconi_etal_2004}.
The statistics of the local dormant SMBHs and distant active QSOs
therefore provide clues to the build-up of the local BH density
across cosmic time, and have implications for the hierarchical
structure formation paradigm, as well as for the interactions
between accreting SMBHs and galaxy bulges.

One of the leading hypotheses for the QSO triggering mechanism is
galaxy mergers \citep[e.g.,][and references
therein]{Hernquist_1989,Carlberg_1990,Kauffmann_Haehnelt_2000,Hopkins_etal_2008a}. In the
hierarchical CDM paradigm, small structures merge to form large
structures, and the merger rate of dark matter halos peaks at
early times, in broad consistency with the observed peak of bright quasar
activities. Gas-rich major mergers (i.e., those with comparable mass ratios)
between galaxies provide an efficient way to channel a large amount
of gas into the central region to trigger starbursts and possibly feed
rapid black hole growth. Observationally this merger hypothesis is
supported by the ULIRG-quasar connection
\citep[e.g.,][]{Sanders_Mirabel_1996,Canalizo_Stockton_2001}, the
signature of recent mergers in quasar hosts
\citep[e.g.,][]{Bennert_etal_2008}, and the small-scale
overdensities of galaxies around luminous quasars or quasar
binaries \citep[e.g.,][]{Fisher_etal_1996,Bahcall_etal_1997, Serber_etal_2006,
Hennawi_etal_2006, Myers_etal_2008,Strand_etal_2008}.
These observations do not necessarily prove that mergers are
directly responsible for triggering QSO activity, but they at least
suggest that QSO activity is coincident with mergers in many
cases.

The last decade has seen a number of models for QSO evolution
based on the merger hypothesis
\citep[e.g.,][]{Haiman_Loeb_1998,Kauffmann_Haehnelt_2000,Wyithe_Loeb_2002,
Wyithe_Loeb_2003,Volonteri_etal_2003,Hopkins_etal_2008a}, which
agree with the bulk of QSO observations reasonably well. Motivated
by the success of these merger-based QSO models, we revisit this
problem in this paper with improved observational data on SMBHs
and QSOs from dedicated large surveys, and with updated knowledge
of the merger rate as inferred from recent numerical simulations.
The present work is different from previous studies in many
aspects in both methodology and the observational tests used. Our
goal is to check if a simple merger-based cosmological QSO
framework can reproduce all the observed statistics of SMBHs (both
active and dormant), and to put constraints on the physical
properties of SMBH growth. Our model is observationally motivated,
therefore we do not restrict ourselves to theoretical predictions
of SMBH/QSO properties from either cosmological or merger event
simulations.

The paper is organized as follows. In \S\ref{subsec:SMBH_demo} we
review some aspects of local SMBH demographics and QSO luminosity
function; in \S\ref{subsec:qso_clustering} we briefly review the
current status of quasar clustering observations;
\S\S\ref{subsec:intro_halo_merger} and
\ref{subsec:intro_gal_merger} discuss the halo and subhalo merger
rate from numerical simulations. Our model formulism is presented
in \S\ref{sec:formalism} and we present our fiducial model and
compare with observations in \S\ref{sec:ref_model}. We discuss
additional aspects of our model in \S\ref{sec:disc} and conclude
in \S\ref{sec:con}.

We use {\em friends-of-friends (fof)} mass with a link length
$b=0.2$ to define the mass of a halo (roughly corresponding to
spherical overdensity mass with $\Delta\sim 200$ times the mean
{\em matter} density), since the fitting formulae for the halo
mass function are nearly universal with this mass definition
\citep[][]{Sheth_Mo_Tormen_2001,Jenkins_etal_2001,Warren_etal_2006,Tinker_etal_2008}.
For simplicity we neglect the slight difference between {\em fof}
mass and virial mass \citep[Appendix \ref{app:B}; and see][for
discussions on different mass definitions]{White_2002}. Throughout
the paper, $L$ always refers to the bolometric luminosity, and we
use subscripts $_B$ or $_X$ to denote $B$-band or X-ray luminosity
when needed. We adopt a flat $\Lambda$CDM cosmology with
$\Omega_0=0.26$, $\Omega_\Lambda=0.74$, $\Omega_b=0.044$, $h=0.7$,
$\sigma_8=0.78$, $n_s=0.95$. We use the
\citet[][]{Eisenstein_Hu_1999} transfer function to compute the
linear power spectrum, and use the fitting formulae for halo
abundance from \citet{Sheth_Tormen_1999} and for halo linear bias
from \citet{Sheth_Mo_Tormen_2001} based on the ellipsoidal
collapse model and tested against numerical simulations.

\subsection{Supermassive Black Hole Demographics}\label{subsec:SMBH_demo}
In the local universe, the dormant BH mass function (BHMF) can be
estimated by combining scaling relations between BH mass and
galaxy bulge properties, such as the $M_{\bullet}-\sigma$ relation
\citep[e.g.,][]{Gebhardt_etal_2000a,Ferrarese_Merritt_2000,Tremaine_etal_2002}
or the $M_{\bullet}-L_{\rm sph}(M_{\rm sph})$ relation
\citep[e.g.,][]{Magorrian_etal_1998,Marconi_Hunt_2003}, with bulge
luminosity or stellar mass/velocity dispersion functions
\citep[e.g.,][]{Salucci_etal_1999,Yu_Tremaine_2002,Marconi_etal_2004,
Shankar_etal_2004,McLure_Dunlop_2004,Yu_Lu_2004,Yu_Lu_2008,
Tundo_etal_2007,Shankar_etal_2009a}. Although some uncertainties
exist on the usage of these scaling relations at both the high and
low BH mass end
\citep[e.g.,][]{Lauer_etal_2007a,Tundo_etal_2007,Hu_2008,Greene_etal_2008,Graham_2008,Graham_Li_2009},
the total BH mass density is estimated to be
$\rho_{\bullet}\approx 4\times 10^5\,M_\odot{\rm Mpc}^{-3}$ with
an uncertainty of a factor $\sim 1.5$
\citep[e.g.][]{Yu_Lu_2008,Shankar_etal_2009a} -- but the shape of
the local BHMF is more uncertain \citep[see the discussion in
][]{Shankar_etal_2009a}.

It is now generally accepted that local SMBHs were once luminous
QSOs
\citep[][]{Salpeter_1964,Zeldovich_Novikov_1964,Lynden-Bell_1969}.
An elegant argument tying the active QSO population to the local
dormant SMBH population was proposed by \citet[][]{Soltan_1982}: if
SMBHs grow mainly through a luminous QSO phase, then the accreted
luminosity density of QSOs to $z=0$ should equal the local BH mass
density:
\begin{equation}\label{eqn:soltan}
\rho_{\rm \bullet,acc}=\int_0^{\infty}\frac{dt}{dz}dz\int_0^\infty
\frac{(1-\epsilon)L}{\epsilon c^2}\Phi(L,z)dL\approx
\rho_{\bullet}\ ,
\end{equation}
where $\Phi(L,z)$ is the bolometric luminosity function (LF) per
$L$ interval. Given the observed QSO luminosity function, a
reasonably good match between $\rho_{\rm \bullet,acc}$ and
$\rho_{\rm \bullet}$ can be achieved if the average radiative
efficiency $\epsilon\sim 0.1$
\citep[e.g.,][]{Yu_Tremaine_2002,Shankar_etal_2004,Marconi_etal_2004,Hopkins_etal_2007a,Shankar_etal_2009a}.
The exact value of $\epsilon$, however, is subject to some
uncertainties from the luminosity function and local BH mass
density determination.

Assuming that SMBH growth is through gas accretion, an extended
version of the Soltan argument can be derived using the continuity
equation\footnote{Note that here the source term, i.e., the
creation of a BH with mass $M_{\bullet}$ by mechanisms other than
gas accretion, is generally neglected
\citep[e.g.,][]{Small_Blandford_1992,Merloni_2004,Shankar_etal_2009a}.
One possible such mechanism is BH coalescence, which modifies the
shape of the BH mass function but does not change the total BH
mass density in the classical case (i.e., neglecting mass loss
from gravitational radiation). Dry mergers at low redshift may act
as a surrogate of shaping the BH mass function.}
\citep[cf.][]{Small_Blandford_1992}:
\begin{equation}\label{eqn:conti}
\frac{\partial n(M_{\bullet},t)}{\partial t}+\frac{\partial
[n(M_\bullet,t)\langle\dot{M}_\bullet\rangle]}{\partial
M_\bullet}=0\ ,
\end{equation}
where $n(M_{\bullet},t)$ is the BH mass function per $dM_\bullet$,
and $\langle\dot{M}_\bullet\rangle$ is the mean accretion rate of
BHs with mass $(M_{\bullet},M_{\bullet}+dM_{\bullet})$ averaged
over both active and inactive BHs. Given the luminosity function,
and a model connecting luminosity to BH mass, one can derive the
BH mass function at all times by integrating the continuity
equation
\citep[e.g.,][]{Marconi_etal_2004,Merloni_2004,Shankar_etal_2009a}.

Since the local BHMF and the QSO luminosity function together
determine the assembly history of the cosmic SMBH population, any
cosmological model of AGN/quasars must first reproduce the
observed luminosity function
\citep[e.g.,][]{Haiman_Loeb_1998,Kauffmann_Haehnelt_2000,
Wyithe_Loeb_2002,Wyithe_Loeb_2003,Volonteri_etal_2003,Lapi_etal_2006,Hopkins_etal_2008a,Shankar_etal_2009a}. This is also one of the central themes of this paper.

There has been great progress in the measurements of the QSO
luminosity function over wide redshift and luminosity ranges for
the past decade, mostly in the optical band
\citep[e.g.,][]{Fan_etal_2001,Fan_etal_2004,Wolf_etal_2003,Croom_etal_2004,
Richards_etal_2005,Richards_etal_2006a,Jiang_etal_2006,Fontanot_etal_2007},
and the soft/hard X-ray band
\citep[e.g.,][]{Ueda_etal_2003,Hasinger_etal_2005,Silverman_etal_2005,Barger_etal_2005}.
While wide-field optical surveys provide the best constraints on
the bright end of the QSO luminosity functions, deep X-ray surveys
can probe the obscured faint end AGN population. Although it has
been known for a while that the spatial density of
optically-selected bright quasars peaks at redshift $z\sim 2-3$,
the spatial density of fainter AGNs selected in the X-ray seems to
peak at lower redshift
\citep[e.g.,][]{Steffen_etal_2003,Ueda_etal_2003,Hasinger_etal_2005},
a trend now confirmed in other observational bands as well
\citep[e.g.,][and references
therein]{Bongiorno_etal_2007,Hopkins_etal_2007a} -- the
manifestation of the so-called {\em downsizing} of the cosmic SMBH
assembly.

\citet{Hopkins_etal_2007a} compiled QSO luminosity
function data from surveys in various bands (optical, X-ray, IR,
etc). Assuming a general AGN spectral energy distribution (SED)
and a column density distribution for obscuration,
\citet{Hopkins_etal_2007a} were able to fit a universal bolometric
luminosity function based on these data. We adopt their compiled
bolometric LF data in our modeling. The advantage of using the
bolometric LF data is that both unobscured and obscured SMBH
growth are counted in the model; but we still suffer from
the systematics of bolometric corrections. The nominal
statistical/systematic uncertainty level of the bolometric LF is
$\sim 20-30\%$ \citep[cf.,][]{Shankar_etal_2009a}.


\subsection{Quasar Clustering}\label{subsec:qso_clustering}

A new ingredient in SMBH studies, due to dedicated large-scale
optical surveys such as the 2QZ \citep{Croom_etal_2004} and SDSS
\citep{SDSS}, is quasar clustering. While quasar clustering studies
can be traced back to more than two decades ago
\citep[e.g.,][]{Shaver_1984}, statistically significant results
only came very recently
\citep[e.g.,][]{PMN_2004,Croom_etal_2005,Porciani_Norberg_2006,
Myers_etal_2006,Myers_etal_2007a,Myers_etal_2007b,Shen_etal_2007b,
da_Angela_2008,Padmanabhan_etal_2008,Shen_etal_2008a,Shen_etal_2009a,Ross_etal_2009}.

Within the biased halo clustering picture
\citep[e.g.,][]{Kaiser_1984,BBKS,Mo_White_1996}, the observed QSO two-point correlation
function implies that
quasars live in massive dark matter halos and are biased tracers
of the underlying dark matter. For optically luminous quasars
($L_{\rm bol}\gtrsim 10^{45}\,{\rm ergs^{-1}}$), the host halo
mass inferred from clustering analysis is a few times
$10^{12}\,h^{-1}M_{\odot}$. Thus quasar clustering provides
independent constraints on how SMBHs occupy dark matter halos,
where the abundance and evolution of the latter population can be
well studied in analytical theories and numerical simulations
\citep[e.g.,][]{Press_Schechter_1974,Bond_etal_1991,Lacey_Cole_1993,
Mo_White_1996,Sheth_Tormen_1999,Sheth_Mo_Tormen_2001,Springel_etal_2005a}.
By comparing the relative abundance of quasars and their host
halos, one can constrain the average quasar duty cycles or
lifetimes
\citep[e.g.,][]{Cole_Kaiser_1989,Martini_Weinberg_2001,Haiman_Hui_2001}
to be $\lesssim 10^8$ yr, which means bright quasars are short
lived.

More useful constraints on quasar models come from studies of
quasar clustering as a function of redshift and luminosity. The redshift evolution
of quasar clustering constrains the evolution of duty cycles of active accretion,
while the luminosity dependence of quasar clustering
puts constraints on quasar light curves (LC).
Successful cosmological quasar models therefore not only need to
account for quasar abundances (LF), but also need to explain
quasar clustering properties, both as function of redshift and
luminosity. We will use quasar clustering observations
as additional tests on our quasar models, as has been done in
recent studies
\citep[e.g.,][]{Hopkins_etal_2008a,Shankar_etal_2009b,Thacker_etal_2009,Bonoli_etal_2009,Croton_2009}.

The luminosity dependence of quasar bias can be modeled as
follows. Assume at redshift $z$ the probability distribution of
host halo mass given bolometric luminosity $L$ is $p(M|L,z)$, then
the halo averaged bias factor is:
\begin{equation}\label{eqn:B_L}
b_L(L,z)=\int b_M(M,z)p(M|L,z)dM\ ,
\end{equation}
where $b_M(M,z)$ is the linear bias factor for halos with mass $M$
at redshift $z$. The derivation of the probability distribution
$p(M|L,z)\equiv dP(M|L,z)/dM$ is described in later sections
[Eqn.\ (\ref{eqn:prob_M_L})].

\subsection{Dark Matter Halo Mergers}\label{subsec:intro_halo_merger}

In the hierarchical universe small density perturbations grow
under gravitational forces and collapse to form virialized halos
(mostly consisting of dark matter). Smaller halos coalesce and
merge into larger ones. Early work on the merger history of dark
matter halos followed the extended Press-Schechter (EPS) theory
\citep[e.g.,][]{Press_Schechter_1974,Bond_etal_1991,Lacey_Cole_1993},
which is based on the spherical collapse model
\citep[][]{Gunn_Gott_1972} with a constant barrier for the
collapse threshold (the critical linear overdensity
$\delta_c\approx 1.68$ is independent on mass, albeit it depends
slightly on cosmology in the $\Lambda$CDM universe). Although the
(unconditional) halo mass function $n(M,z)$ predicted by the EPS
theory agrees with numerical $N$-body simulations reasonably well,
it is well known that it overpredicts (underpredicts) the halo
abundance at the low (high) mass end \citep[][and references
therein]{Sheth_Tormen_1999}. Motivated by the fact that halo
collapses are generally triaxial, \citet{Sheth_Tormen_1999}, based
on earlier work by \citet{Bond_Myers_1996}, proposed the
ellipsoidal collapse model with a moving barrier where the
collapse threshold also depends on mass. By imposing this mass
dependence on collapse barrier, the abundance of small halos is
suppressed and fitting formulae for the (unconditional) halo mass
function are obtained, which agree with $N$-body simulations much
better than the spherical EPS predictions
\citep[e.g.,][]{Sheth_Tormen_1999,Sheth_Mo_Tormen_2001,Jenkins_etal_2001,Warren_etal_2006,Tinker_etal_2008}.

In addition to the {\em unconditional} halo mass function, the
{\em conditional} mass function $n(M_1,z_1|M_0,z_0)$ gives the
mass spectrum of progenitor halos at an earlier redshift $z_1$ of
a descendant halo $M_0$ at redshift $z_0$. This conditional mass
function thus contains information about the merger history of
individual halos and can be used to generate halo merger trees. A
simple analytical form for $n(M_1,z_1|M_0,z_0)$ can be obtained in
the spherical EPS framework \citep[e.g.,][]{Lacey_Cole_1993}; for
the ellipsoidal collapse model, an exact but
computationally consuming solution of the conditional mass
function can be obtained by solving the integral equation proposed
by \citet{zhang_Hui_2006}. Recently, \citet[]{zhang_etal_2008a}
derived analytical formulae for $n(M_1,z_1|M_0,z_0)$ in the limit
of small look-back times for the ellipsoidal collapse model.

Alternatively, the halo merger rate can be retrieved directly from
large volume, high resolution cosmological $N$-body simulations,
which bypasses the inconsistencies between the spherical EPS
theory and numerical simulations. Using the product of the
Millennium Simulation \citep{Springel_etal_2005a},
\citet{Fakhouri_Ma_2008} quantified the mean halo merger rate per
halo for a wide range of descendant (at $z=0$) halo mass
$10^{12}\lesssim M_0\lesssim 10^{15}\,M_\odot$, progenitor mass
ratio $10^{-3}\lesssim \xi\le 1$ and redshift $0\le z\lesssim 6$,
and they provided an almost universal fitting function for the
mean halo merger rate to an accuracy $\lesssim 20\%$ within the
numerical simulation results:
\begin{eqnarray}\label{eqn:halo_merger_rate}
\frac{B(M_0,\xi,z)}{n(M_0,z)}=0.0289\bigg(\frac{M_{\rm 0}}{1.2\times10^{12}\,M_\odot}\bigg)^{0.083}\xi^{-2.01}\times \nonumber \\
\exp\bigg[\bigg(\frac{\xi}{0.098}\bigg)^{0.409}\bigg]\bigg(\frac{d\delta_{c,z}}{dz}\bigg)^{0.371}\
.
\end{eqnarray}
Here $B(M_0,\xi,z)$ is the instantaneous merger rate at redshift
$z$, for halos with mass $(M_0,M_0+dM_0)$ and with progenitor mass
ratio in the range $(\xi, \xi+d\xi)$ [$B(M_0,\xi,z)$ is in units
of number of mergers $\times {\rm
Mpc}^{-3}M_\odot^{-1}dz^{-1}d\xi^{-1}$], $n(M_0,z)$ is the halo
mass function (in units of ${\rm Mpc}^{-3}{M_\odot}^{-1}$),
$\xi\equiv M_2/M_1\le 1$ where $M_1\ge M_2$ are the masses of the
two progenitors, and $\delta_{c,z}=\delta_c/D(z)$ is the linear
density threshold for spherical collapse with $D(z)$ the linear
growth factor. We adopt Eqn.\ (\ref{eqn:halo_merger_rate}) to
estimate the halo merger rate in our modeling. The mass definition
used here is {\em friends-of-friends} mass $M_{\rm fof}$ with a
link length $b=0.2$. For the range of mass ratios we are
interested in (i.e., major mergers), the halo merger rate derived
from the spherical EPS model can overpredict the merger rate by up
to a factor of $\sim 2$ \citep[][]{Fakhouri_Ma_2008}.

\subsection{Galaxy (subhalo) Mergers}\label{subsec:intro_gal_merger}


Galaxies reside in the central region of dark matter halos where
the potential well is deep and gas can cool to form stars. When a
secondary halo merges with a host halo, it (along with its central
galaxy) becomes a satellite within the virial radius of the host
halo. It will take a dynamical friction time for the subhalo to
sink to the center of the primary halo, where the two galaxies
merge. The subsequent galaxy (subhalo) merger rate is therefore
different from the halo merger rate discussed in
\S\ref{subsec:intro_halo_merger}, and a full treatment with all
the dynamics (tidal stripping and gravitational shocking of
subhalos) and baryonic physics is rather complicated.

The simplest argument for the galaxy (subhalo) merger timescale is
given by dynamical friction \citep[]{Chandra_1943,BT_1987}:
\begin{equation}\label{eqn:df}
\tau_{\rm df}\approx \frac{f_{\rm df}\Theta_{\rm
orb}}{\ln\Lambda}\frac{M_{\rm host}}{M_{\rm sat}}\tau_{\rm dyn}\ ,
\end{equation}
where $M_{\rm host}$ and $M_{\rm sat}$ are the masses for the host
and satellite halos respectively, $\ln\Lambda\approx \ln (M_{\rm
host}/M_{\rm sat})$ is the Coulomb logarithm, $\Theta_{\rm orb}$
is a function of the orbital energy and angular momentum of the
satellite, $f_{\rm df}$ is an adjustable parameter and $\tau_{\rm
dyn}\equiv r/V_c(r)$ is the halo dynamical timescale, usually
estimated at the halo virial radius $r_{\rm vir}$. This formula
(\ref{eqn:df}) is valid at the small satellite mass limit $M_{\rm
host}/M_{\rm sat}\gg 1$; although it is also used in cases of
$M_{\rm host}\gtrsim M_{\rm sat}$ in many semi-analytical models
(SAMs) with the replacement of $\ln\Lambda =(1/2)\ln[1+(M_{\rm
host}/M_{\rm sat})^2]$ (i.e., the original definition of the
Coulomb logarithm) or $\ln\Lambda=\ln(1+M_{\rm host}/M_{\rm
sat})$. However, in recent years deviations from the predictions
by Eqn.\ (\ref{eqn:df}) in numerical simulations have been
reported for both the $M_{\rm sat}\ll M_{\rm host}$ and the
$M_{\rm sat}\lesssim M_{\rm host}$ regimes
\citep[e.g.,][]{taffoni_etal_2003,monaco_etal_2007,Boylan-Kolchin_etal_2008,Jiang_etal_2008}.
In particular even in the regime $M_{\rm sat}/M_{\rm host}\ll 1$
where the original Chandrasekhar formula is supposed to work,
Eqn.\ (\ref{eqn:df}) substantially underestimates the merger
timescale from simulations by a factor of a few (getting worse at
lower mass ratios). This is because Eqn.\ (\ref{eqn:df}) is
derived by treating the satellite as a rigid body, while in
reality the satellite halo loses mass via tidal
stripping\footnote{The galaxy associated with the subhalo, on the
other hand, does not suffer from mass stripping since it sits in
the core region of the subhalo. } and hence the duration of
dynamical friction is greatly extended.

Given the limitations of analytical treatments, we therefore
retreat to numerical simulation results. Fitting formulae for the
galaxy merger timescales within merged halos have been provided by
several groups
\citep[][]{taffoni_etal_2003,monaco_etal_2007,Boylan-Kolchin_etal_2008,Jiang_etal_2008},
and the merger rate of subhalos has also been directly measured
from simulations \citep[][]{Wetzel_etal_2008a,Stewart_etal_2008}.

\begin{figure}
  \centering
    \includegraphics[width=0.48\textwidth]{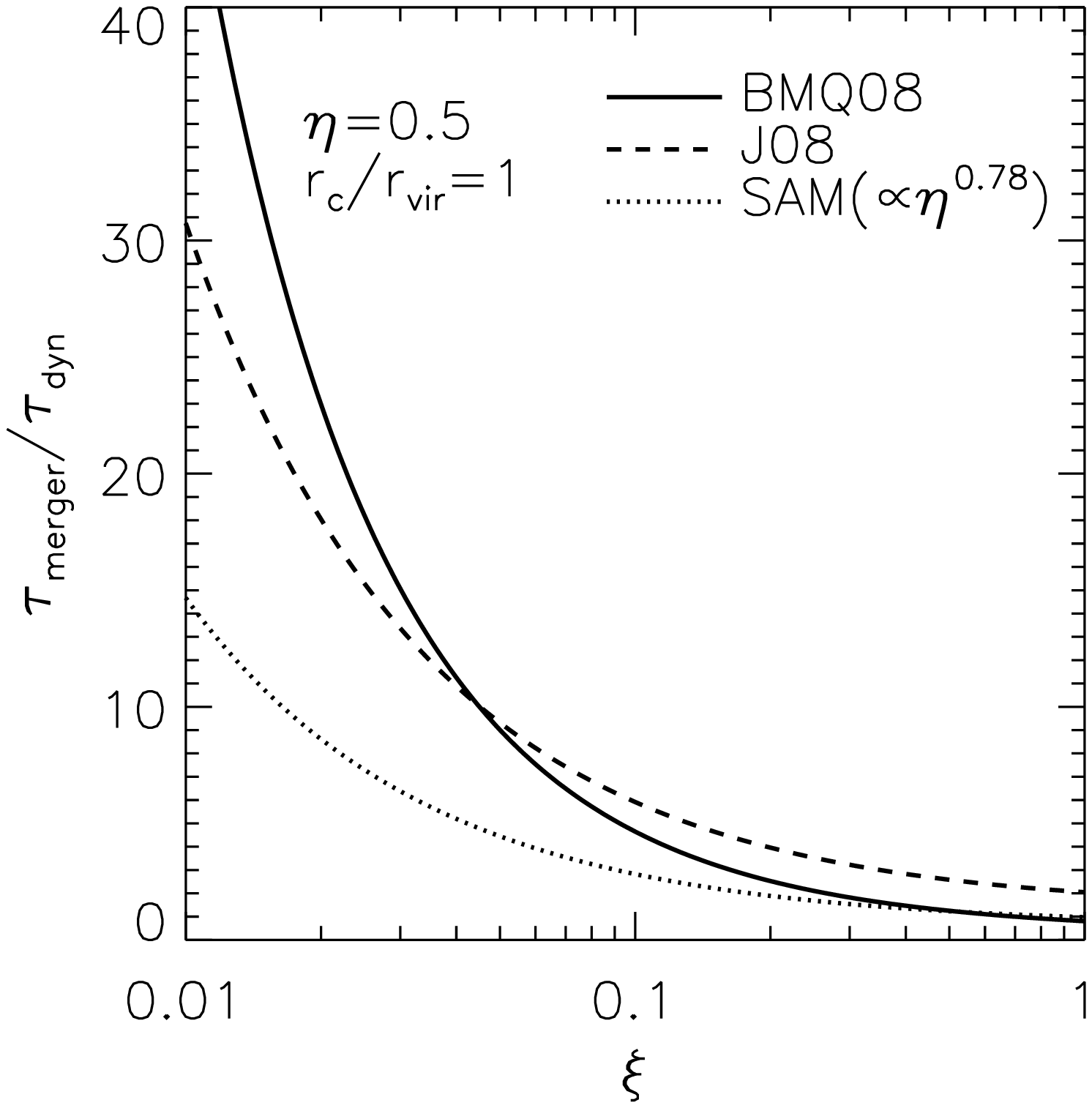}
    \includegraphics[width=0.48\textwidth]{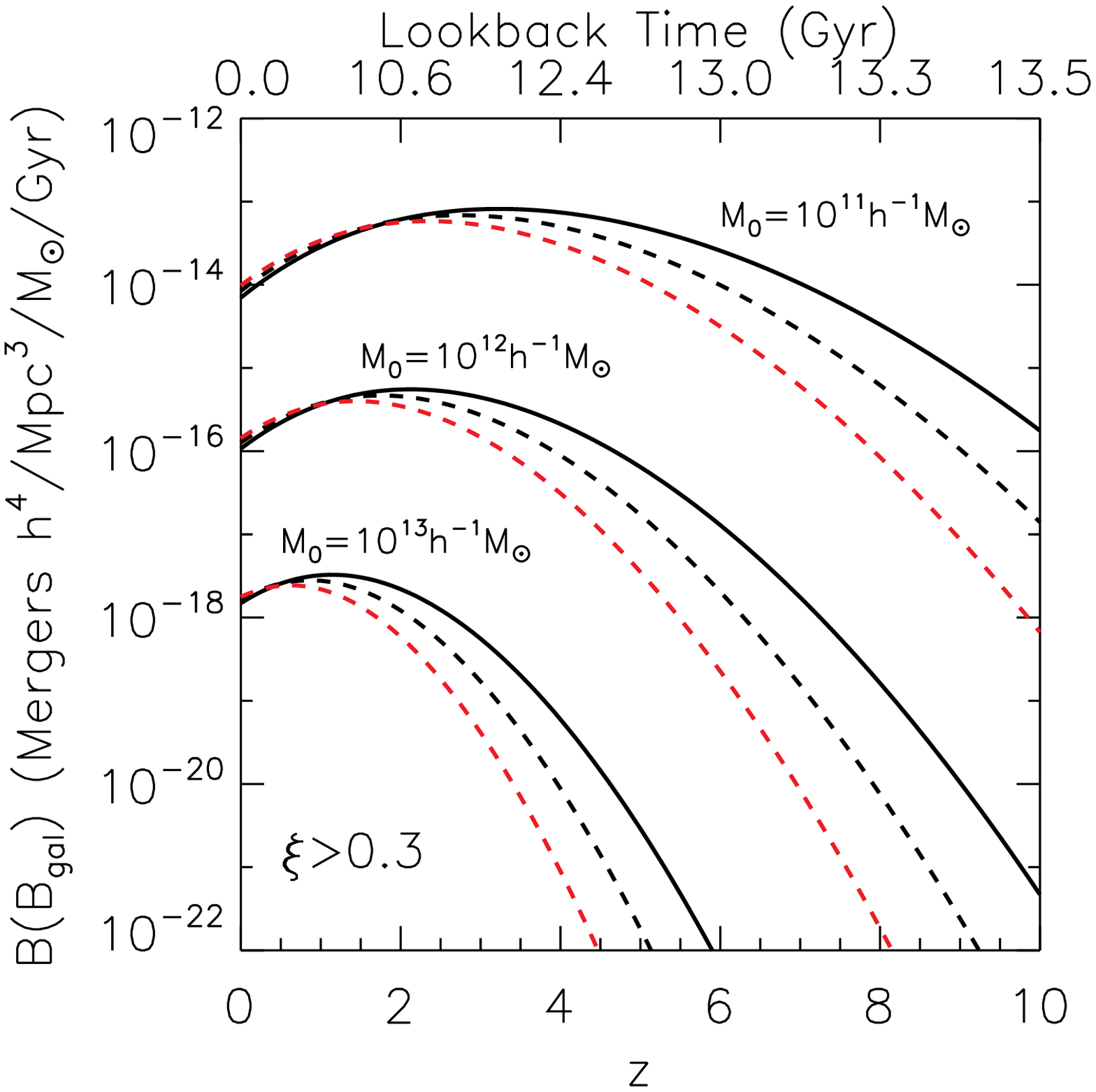}
    \caption{Merger timescales and (sub)halo merger rates. {\em Upper}: comparison of the subhalo merger
    timescale from three dynamical friction prescriptions. {\em Bottom}: the redshift
    evolution of the halo merger rate (black solid lines), and the subhalo merger rate using the
    \citet{Boylan-Kolchin_etal_2008} (black dashed lines),
    and \citet{Jiang_etal_2008} (red dashed lines) prescriptions for the merger
    timescale, for three (post)merger halo masses.}
    \label{fig:tau_merger}
\end{figure}

The fitting formula for $\tau_{\rm merger}$ in
\citet{Jiang_etal_2008}, who used hydro/$N$-body simulations, has
the following form:
\begin{equation}\label{eqn:sub_merger_rate}
\tau_{\rm merger}(\xi,z)=\frac{0.94\eta^{0.6}+0.6}{2\times
0.43}\frac{1}{\xi\ln[1+(1/\xi)]}\frac{r_{\rm vir}}{V_c}\ ,
\end{equation}
where $\eta$ is the circularity parameter [$\eta=(1-e^2)^{1/2}$,
and $e$ is the orbit eccentricity]. In deriving this equation
Jiang et al.\ have removed the energy dependence of the orbit, i.e.,
they fix $r_{\rm c}=r_{\rm vir}$ where $r_{\rm c}$ is the circular
orbit that has the same energy as the satellite's orbit. In their simulation
the ratio of $r_{\rm c}/r_{\rm vir}$ ranges from $\sim 0.6-1.5$ and has a
median value $\sim 1$. The distribution of circularity $\eta$ in their
simulation (see their eqn. 7) has a median value $\sim 0.5$, so in
what follows we take $\eta=0.5$. The redshift dependence of
$\tau_{\rm merger}$ is thus the same as that of the halo dynamical time
$\tau_{\rm dyn}=r_{\rm vir}/V_c\propto [\Delta_{\rm
vir}(z)]^{-1/2}(1+z)^{-3/2}$ (see Appendix \ref{app:B}).

Alternatively, the fitting formula of $\tau_{\rm merger}$ in
\citet{Boylan-Kolchin_etal_2008} is given by:
\begin{equation}\label{eqn:sub_merger_rate2}
\tau_{\rm
merger}(\xi,z)=0.216\frac{\exp(1.9\eta)}{\xi^{1.3}\ln[1+(1/\xi)]}\frac{r_{\rm
c}(E)}{r_{\rm vir}}\tau_{\rm dyn}\ .
\end{equation}

Combining the halo merger rate (\ref{eqn:halo_merger_rate}) and
galaxy (subhalo) merger timescale (\ref{eqn:sub_merger_rate}) or
(\ref{eqn:sub_merger_rate2}) we derive the instantaneous merger
rate of galaxies within a merged halo of mass $M_0$ for progenitor
mass ratio $\xi$ and at redshift $z$:
\begin{equation}\label{eqn:gal_merger_rate}
B_{\rm gal}(M_0,\xi,z)=B[M_0,\xi,z_{\rm e}(z,\xi)]\frac{dz_{\rm
e}}{dz}\ ,
\end{equation}
where $z_{\rm e}(z,\xi)$ is a function of $(z,\xi)$ and is
determined by:
\begin{equation}
t_{\rm age}(z)-t_{\rm age}(z_{\rm e})=\tau_{\rm merger}(\xi,z_{\rm
e})\ ,
\end{equation}
where $t_{\rm age}(z)$ is the cosmic time at redshift $z$. Again, here
$B_{\rm gal}(M_0,\xi,z)$ is the number of mergers per volume per
halo mass per mass ratio per redshift. We find
that $dz_{\rm e}/dz$ is almost constant {\em at all redshifts} for
$\xi=0.1-1$ and monotonically decreases as $\xi$ increases. This
constant can be approximated by its asymptotic value at large $z$
when $\Omega(z)\rightarrow 1$:
\begin{equation}\label{eqn:dze_dz1}
\frac{dz_{\rm e}}{dz}\approx \bigg
\{1+0.22\bigg[\xi\ln(1+1/\xi)\bigg]^{-1}\bigg \}^{2/3}\ ,
\end{equation}
for the fitting formula of $\tau_{\rm merger}$ in
\citet{Jiang_etal_2008}, or
\begin{equation}\label{eqn:dze_dz2}
\frac{dz_{\rm e}}{dz}\approx \bigg
\{1+0.09\bigg[\xi^{1.3}\ln(1+1/\xi)\bigg]^{-1}\bigg \}^{2/3}\ ,
\end{equation}
for the fitting formula of $\tau_{\rm merger}$ in
\citet{Boylan-Kolchin_etal_2008}.

Unfortunately, current studies on the galaxy merger timescale
have not converged yet, and different groups do report similar but
quantitatively different results, at least partly caused by the
different definitions and setups in their numerical simulations
(e.g., hydro/$N$-body versus pure $N$-body simulations, halo
finding algorithms and definition of mergers, etc). These fitting
formulae are generally good within a factor of $\sim 2$
uncertainty. This uncertainty in the galaxy merger timescale
within DM halos will lead to quite substantial differences in the
galaxy merger rate at high redshift. To see this, we show in Fig.\
\ref{fig:tau_merger} the comparison of different fitting formulae
for $\tau_{\rm merger}$ and their consequences. In the upper panel
of Fig.\ \ref{fig:tau_merger} we show the ratio of $\tau_{\rm
merger}/\tau_{\rm dyn}$ for the most likely orbit with circularity
$\eta=0.5$ and $r_{\rm c}=r_{\rm vir}$ using the fitting formula
in \citet[][]{Boylan-Kolchin_etal_2008} (solid line),
\citet[][]{Jiang_etal_2008} (dashed line), and a commonly-used SAM
prescription: $\tau_{\rm
merger}=1.16\eta^{0.78}[\xi\ln(1+1/\xi)]^{-1}r_{\rm c}/r_{\rm
vir}$ (dotted line). For the major merger regime $\xi\gtrsim 0.3$
the fitting formula in \citet{Boylan-Kolchin_etal_2008} agrees
with the SAM prescription well, and they both approach the
dynamical time at the high mass ratio end. However, the fitting
formula in \citet{Jiang_etal_2008} yields a factor of $\sim 2$
longer than dynamical time at the high mass ratio end.

In the lower panel of Fig.\ \ref{fig:tau_merger} we show the halo
and galaxy major merger rates {\em per unit time} (integrated over
$\xi>0.3$), as function of redshift. The black solid lines show
the halo major merger rate, the black dashed lines show the galaxy
merger rate using the fitting formula of $\tau_{\rm merger}$ from
\citet{Boylan-Kolchin_etal_2008}, and the red dashed lines show
the galaxy merger rate adopting the fitting formula from
\citet{Jiang_etal_2008}. In general the galaxy merger rates fall
below the halo merger rate at high redshift and take over at lower
redshift. At redshift $z>4$, the galaxy merger rate using the
\citet{Jiang_etal_2008} formula is lower by almost an order of
magnitude than that using the \citet{Boylan-Kolchin_etal_2008}
formula (as well as the SAM prescription, since it agrees with
eqn.\ \ref{eqn:sub_merger_rate2} for $\xi>0.3$), which makes it
difficult to produce the quasar abundance at high redshift (see
later sections).

It is worth noting that although the merger rate peaks at some
redshift, this trend is {\em hierarchical} such that it peaks at
later time for more massive halos. This is somewhat in
contradiction to the {\em downsizing} of the QSO luminosity
function. This apparent discrepancy is likely caused by the fact
that at low reshift, it becomes progressively more difficult for
the black hole to accrete efficiently in massive halos because of
the global deficit of cold gas due to previous mergers experienced
by massive halos and/or possible feedback quenches
\citep[e.g.,][]{Kauffmann_Haehnelt_2000,Croton_etal_2006}. We will
treat this fraction of QSO-triggering merger event as function of
redshift and halo mass explicitly in our modeling.

If we choose to normalize the total merger rate $B$ ($B_{\rm
gal}$) by $n(M_0,z)$, the abundance of descendant halos with mass
$M_0$, at the same redshift $z$ for halo mergers and galaxy
mergers, then $B_{\rm gal}/n$ falls below $B/n$ at high redshift,
then catches up and exceeds $B/n$ a little, and evolves more or
less parallel to $B/n$ afterwards. Therefore the evolution in
$B_{\rm gal}/n$ is shallower than that in $B/n$, consistent with
the findings by \citet{Wetzel_etal_2008a} once the different
definitions of $B/n$ and halo mass are taken into account.

To summarize \S\ref{subsec:intro_halo_merger} and
\S\ref{subsec:intro_gal_merger}, we have compared the halo merger
rate and galaxy merger rate as functions of redshift, (post)merger
halo mass and mass ratio with different prescriptions for the
dynamical friction timescale. It is, however, unclear on which
rate we should link to the QSO triggering rate. It is true that it
will take a dynamical friction time for the two galaxies to merge
after their host halo merged. But the black hole accretion might
be triggered very early during their first orbital crossing, well
before the galaxies merge. In what follows, we adopt the halo
merger rate to model the QSO triggering rate, and we discuss the
consequences of adopting the alternative subhalo merger rates in
\S\ref{subsec:disc1}.


\begin{deluxetable*}{lll}
\tabletypesize{\footnotesize}  
\tablecolumns{3} \tablewidth{0.9\textwidth} \tablecaption{Notation
and Model Parameters \label{table:notation}} \tablehead{Symbol &
Description & Value/Units}
\startdata $\xi$\dotfill & Halo mass ratio & $0<\xi\le 1$\\
$M_0$\dotfill & Postmerger halo mass & \\
$M_{\bullet}$\dotfill & Black hole mass & \\
$B,B_{\rm gal}$\dotfill & Halo(subhalo) merger rate & ${\rm Mpc}^{-3}M_\odot^{-1}dz^{-1}d\xi^{-1}$ \\
$\Phi(L,z)$\dotfill & QSO luminosity function & ${\rm Mpc}^{-3}dL^{-1}$ \\
$B_{L}$\dotfill  & QSO triggering rate & ${\rm Mpc}^{-3}dL^{-1}dz^{-1}$ \\
$L_{\rm peak}$\dotfill & QSO peak luminosity &\\
$z^\prime$\dotfill & QSO triggering redshift & \\
$f_{\rm QSO}(z,M_0)$\dotfill & QSO triggering fraction & \\
\hline\\
&{\bf The $L_{\rm peak}-M_0$ relation} &  \\
\hline\\
$\gamma$\dotfill & Slope & $5/3$\\
$C(z=0)$\dotfill & Normalization at $z=0$& $\log(6\times10^{45})-12\gamma$ \\
$\sigma_L$\dotfill & Scatter& $0.28$ dex \\
$C(z)=C(z=0)+\beta_1\log(1+z)$\dotfill &Evolution in normalization
&
$\beta_1=0.2\gamma=1/3$\\
\hline\\
&{\bf The light curve model}  &\\
\hline\\
$\epsilon$\dotfill & Radiative efficiency & $0.1$ \\
$l$\dotfill & Eddington luminosity per $M_\odot$ &
$1.26\times10^{38}\,{\rm ergs^{-1}}M_\odot^{-1}$\\
$t_{\rm Salpeter}$\dotfill & $e$-folding time & $\displaystyle\frac{\epsilon c^2}{l(1-\epsilon)\lambda_0}$\\
$f$ & Fraction of seed BH mass to peak BH mass & $10^{-3}$\\
$t_{\rm peak}$\dotfill & Time to reach the peak luminosity & $(-\ln f)t_{\rm Salpeter}$ \\
$\lambda_0$\dotfill & Eddington ratio before $t_{\rm peak}$ & $3$ \\
$\alpha$\dotfill & Power-law slope of the decaying phase & $2.5$ \\
\hline\\
&{\bf The QSO-triggering rate}  &\\
\hline\\
$\xi_{\rm min}$\dotfill & Minimum mass ratio & 0.25\\
$M_{\rm min}$\dotfill  & Exponential lower mass cut & $3\times
10^{11}\,h^{-1}M_\odot$\\
$M_{\rm max}(z)=M_{\rm quench}(1+z)^\beta$\dotfill & Exponential
upper
mass cut & $10^{12}(1+z)^{3/2}\,h^{-1}M_\odot$\\
\enddata
\tablecomments{Parameter values are for the fiducial model.}
\end{deluxetable*}

\subsection{BH Fueling During Mergers}\label{subsec:BH_merger}

Modeling black hole fueling during mergers from first
principles is not trivial: aside from the lack of physical inputs,
most hydrodynamical simulations do not yet have the dynamical range to even
resolve the outer Bondi radius. Although semi-analytical
treatments of BH accretion are possible
\citep[e.g.,][]{Granato_etal_2004,monaco_etal_2007}, we do not
consider such treatments in this paper because of the poorly
understood accretion physics.

Throughout this paper we adopt the following {\em ansatz}:

\begin{itemize}
\item {\em QSO activity is triggered by a gas-rich major merger
event ($\xi_{\rm min}<\xi\le 1$).}

\item {\em The build-up of the cosmic SMBH population is mainly
through gas accretion during the QSO phase.}

\item {\em The QSO light curve follows a universal form $L(L_{\rm
peak},t)$, where the peak luminosity $L_{\rm peak}$ is correlated
with the mass of the merged halo $M_0$.}

\end{itemize}

Once we have specified the QSO light curve model, and estimated
the QSO triggering rate from the major merger rate, we can convolve
them to derive the QSO luminosity function. Within this evolutionary
QSO framework, we can derive the distributions of host halo masses
and BH masses for any given instantaneous luminosity and redshift, allowing
us to compare with observations of quasar clustering and Eddington
ratio distributions. The details of this framework are elaborated in \S\ref{sec:formalism}.

We note that, any correlations established with our model are only
for hosts where a gas-rich major merger is triggered and
self-regulated BH growth occurs. Host halos which experience many
minor mergers or dry mergers may deviate from these relations. We
will come back to this point in the discussion section.

\section{Model Formalism}\label{sec:formalism}
In this section we describe our model setup in detail. Some
notations and model parameters are summarized in Table
\ref{table:notation} for clarity.

\subsection{Determining the Luminosity Function}

Major mergers are generally defined as events with mass ratio
$\xi_{\rm min}\equiv 0.3\lesssim\xi\le 1$, but our framework can be easily generalized to
other values of $\xi_{\rm min}$. Because it takes more than just a major
merger event to trigger QSO activity, we shall model the QSO
triggering rate as a redshift and mass-dependent fraction of halo
merger rate. 
The QSO luminosity function at a given redshift $z$ can be
obtained by combining the triggering rate and the evolution of QSO
luminosity\footnote{Note that we have implicitly assumed that a
second QSO-triggering merger event does not occur during the major
accretion phase of the QSO -- a reasonable assumption since
statistically speaking bright QSOs are short-lived ($t_{\rm
QSO}\ll t_{H}$). Observationally binary/multiple quasars within a
single halo with comparable luminosities are rare occurrences
\citep[$f_{\rm binary}\lesssim 0.1\%$, e.g.,
][]{Hennawi_etal_2006,Hennawi_etal_2009,Myers_etal_2008}, further
supporting this assumption.}:
\begin{equation}\label{eqn:LF_merger}
\Phi(L,z)dL=\int_\infty^z B_{L}[L_{\rm
peak}(L,t_z-t_{z^\prime}),z^\prime]dL_{\rm peak}dz^\prime\ ,
\end{equation}
where $\Phi(L,z)dL$ is the QSO number density in the luminosity
range $(L,L+dL)$ at redshift $z$; $B_{L}(L_{\rm peak},z^\prime)$
is the QSO-triggering rate (merger number density per redshift per
peak luminosity) at redshift $z^\prime$ with peak bolometric
luminosity $L_{\rm peak}(L,t_z-t_{z^\prime})$, which is determined
by the specific form of the light curve; $t_z$ and $t_{z^\prime}$
are the cosmic time at $z$ and $z^\prime$. In integrating this
equation we impose an upper limit of redshift $z_{\rm max}=20$,
but we found that the integral converges well below this redshift
since the merger rate decays rapidly at high redshift. Eqn.\
(\ref{eqn:LF_merger}) implies that the distribution of triggering
redshift $z^\prime$ given $L$ at redshift $z$ is:
\begin{equation}\label{eqn:prob_Lpeak_L}
\frac{dP(L_{\rm peak},z^{\prime}|L,z)}{dz^\prime}=p(L_{\rm
peak},z^{\prime}|L,z)\propto B_L(L_{\rm
peak},z^\prime)\frac{dL_{\rm peak}}{dL}\ ,
\end{equation}
where again $L_{\rm peak}$ is tied to $L$ via the light curve
model. Obviously only quasars with $L_{\rm peak}\ge L$ can
contribute to $\Phi(L,z)$.

The QSO-triggering rate $B_{L}(L_{\rm peak},z)$ is related to the
halo merger rate $B(M_0,\xi,z)$ by:
\begin{equation}\label{eqn:agn_rate}
B_{L}(L_{\rm peak},z)dL_{\rm peak}=f_{\rm
QSO}(z,M_0)\int_{\xi_{\rm min}}^1 B(M_0,\xi,z)d\xi dM_0\ ,
\end{equation}
where we parameterize the fraction $f_{\rm QSO}$ as
\begin{equation}
f_{\rm QSO}(z,M_0)={\cal F}(z)\exp\bigg[-\frac{M_{\rm
min}(z)}{M_0}-\frac{M_0}{M_{\rm max}(z)}\bigg]\ ,
\end{equation}
where ${\cal F}(z)$ describes how $f_{\rm QSO}(z,M_0)$ decreases
towards lower redshift, i.e., the overall reduction due to the
consumption of cold gas. We also introduce exponential cutoffs of
$f_{\rm QSO}$ at both high and low mass such that at each
redshift, halos with too small $M_0$ cannot trigger QSO activity;
on the other hand, overly massive halos cannot cool gas
efficiently and BH growth halts, and the gas-rich fraction may
become progressively smaller at higher halo masses (especially at
low redshift). We discuss choices for these parameters later in
\S\ref{sec:ref_model}.


To proceed further we must specify the relation between $L_{\rm
peak}$ and $M_0$, and choose a light curve model. We assume that
the $L_{\rm peak}-M_0$ correlation is a power-law with log-normal
scatter, as motivated by some analytical arguments and
hydrodynamical simulations
\citep[e.g.,][]{Wyithe_Loeb_2002,Wyithe_Loeb_2003,Springel_etal_2005a,Hopkins_etal_2005a,Lidz_etal_2006}:
\begin{eqnarray}\label{eqn_prob_M0_Lpeak}
&&\frac{dP(L_{\rm peak}|M_0)}{d\log L_{\rm peak}}=p(L_{\rm
peak}|M_0)\nonumber\\
&&=(2\pi\sigma_L^2)^{-1/2}\exp\bigg\{-\frac{[\log L_{\rm peak}-(C
+ \gamma\log M_{0})]^2}{2\sigma_L^2}\bigg\}\ ,
\end{eqnarray}
where the mean relation is:
\begin{equation}\label{eqn:Lpeak_M0}
\left\langle\log \left(\frac{L_{\rm peak}}{{\rm erg\,
s^{-1}}}\right)\right\rangle=C + \gamma\log
\left(\frac{M_{0}}{h^{-1}M_\odot}\right)\ ,
\end{equation}
with normalization $C$ and power-law slope $\gamma$. The log
scatter around this mean relation is denoted as $\sigma_L$. When
the scatter between $L_{\rm peak}$ and $M_0$ is incorporated,
Eqn.\ (\ref{eqn:agn_rate}) becomes:
\begin{eqnarray}\label{eqn:agn_rate2}
&&B_{L}(L_{\rm peak},z) \nonumber\\
&&=\int\bigg[ f_{\rm QSO}\int_{\xi_{\rm min}}^1
B(M_0,\xi,z)d\xi\bigg]\frac{M_0}{L_{\rm peak}} p(L_{\rm
peak}|M_0)d\log M_0\ , \nonumber\\
\end{eqnarray}
from which we obtain the probability distribution of postmerger
halo mass $M_0$ at fixed peak luminosity $L_{\rm peak}$ and
redshift $z$:
\begin{eqnarray}\label{eqn:prob_M_Lpeak}
&&\frac{dP(M_0|L_{\rm peak},z)}{d\log M_0}=p(M_0|L_{\rm peak},z)\nonumber\\
&&\propto \bigg[f_{\rm QSO}\int_{\xi_{\rm min}}^1
B(M_0,\xi,z)d\xi\bigg]\frac{M_0}{L_{\rm peak}} p(L_{\rm
peak}|M_0)\ .
\end{eqnarray}
We derive the probability distribution of
host halo mass at given instantaneous luminosity and redshift,
$p(M_0|L,z)$, as:
\begin{eqnarray}\label{eqn:prob_M_L}
&&\frac{dP(M_0|L,z)}{d\log M_0}=p(M_0|L,z)\nonumber\\
&&=\int\frac{dP(M_0|L_{\rm peak},z^\prime)}{d\log
M_0}\frac{dP(L_{\rm peak},z^\prime|L,z)}{dz^\prime}dz^\prime\ ,
\end{eqnarray}
which can be convolved with the halo linear bias $b_M(M_0,z)$ to derive the
QSO linear bias $b_L(L,z)$ at instantaneous luminosity $L$ and redshift $z$.

Since we have assumed that a second QSO-triggering merger event
has not occurred, at redshift $z$ the postmerger halo $M_0$ should
maintain most of its identity as it formed sometime earlier, i.e.,
we neglect mass added to the halo via minor mergers or accretion of diffuse
dark matter between the halo formation time and the time when the
QSO is observed at redshift $z$.

Given the halo mass distribution (\ref{eqn:prob_M_L}), we can
further derive the ``active'' halo mass function hosting a QSO
with $L>L_{\rm min}$:
\begin{equation}\label{eqn:active_hmf}
\frac{d\Psi_{M_0}}{d\log M_0}=\int_{L_{\rm min}}^\infty
\Phi(L,z)dL\frac{dP(M_0|L,z)}{d\log M_0}\ ,
\end{equation}
which can be used to compute the halo duty cycles.

\subsection{The Light Curve Model}

For the light curve model there are several common choices in
merger-based cosmological QSO models
\citep[e.g.,][]{Kauffmann_Haehnelt_2000,Wyithe_Loeb_2002,Wyithe_Loeb_2003}:
a) a light bulb model in which the QSO shines at a constant value $L_{\rm peak}$
for a fixed time $t_{\rm QSO}$; b) an exponential decay model
$L=L_{\rm peak}\exp(-t/t_{\rm QSO})$; c) a power-law decay model
$L=L_{\rm peak}(1+t/t_{\rm QSO})^{-\alpha}$ with $\alpha>0$. Once
a light curve model is chosen, the total accreted mass during the
QSO phase is simply\footnote{Throughout the paper we assume
constant radiative efficiency $\epsilon$. At the very late stage
of evolution or under certain circumstance (i.e., hot gas
accretion within a massive halo), a SMBH may accrete via {\em
radiatively-inefficient accretion flows} (RIAFs) with very low
$\epsilon$ and mass accretion rate \citep[e.g.,][]{Narayan_Yi_1995}. We neglect this possible
accretion state since the mass accreted during this state is most
likely negligible \citep[e.g.][]{Hopkins_etal_2006a}.}:
\begin{equation}
M_{\rm \bullet,relic}=\frac{1-\epsilon}{\epsilon c^2}\int_0^\infty
L(L_{\rm peak},t)dt\ .
\end{equation}

The evolution of the {\em luminosity} Eddington ratio
$\lambda\equiv L/L_{\rm Edd}$ is (neglecting the seed BH mass):
\begin{equation}
\lambda(t)=\frac{L(L_{\rm peak},t)}{lM_\bullet(t)}=\frac{L(L_{\rm
peak},t)}{\displaystyle\frac{l(1-\epsilon)}{\epsilon c^2}\int_0^t
L(L_{\rm peak},t^\prime)dt^\prime }\ ,
\end{equation}
where $l\equiv 1.26\times 10^{38}\,{\rm erg\,s^{-1}}M_\odot^{-1}$
is the Eddington luminosity per $M_\odot$.

Unfortunately none of the three LC models is self-consistent
without having $\lambda\gg 1$ at the very beginning of BH growth,
simply because they neglect the rising part of the light curve. To
remedy this, we must model the evolution of luminosity and BH
growth self-consistently
\citep[e.g.,][]{Small_Blandford_1992,Yu_Lu_2004,Yu_Lu_2008}. We
consider a general form of light curve where the BH first grows
exponentially at constant {\em luminosity} Eddington ratio
$\lambda_0$ \citep[e.g.,][]{Salpeter_1964} to $L_{\rm peak}$ at
$t=t_{\rm peak}$, and then the luminosity decays monotonically as
a power-law \citep[e.g.,][]{Yu_Lu_2008}:
\begin{eqnarray}\label{eqn:LC_model}
L(L_{\rm peak},t)  & = &
 \left\{
\begin{array}{lc}
L_{\rm peak}\displaystyle\exp\bigg[\frac{l(1-\epsilon)\lambda_0}{\epsilon c^2}(t-t_{\rm peak})\bigg], & 0\le t\le t_{\rm peak} \\ \\
L_{\rm peak}\displaystyle\bigg(\frac{t}{t_{\rm
peak}}\bigg)^{-\alpha}, & t\ge t_{\rm peak}\ ,
\end{array} \right.
\end{eqnarray}

where $t_{\rm peak}$ is determined by:
\begin{equation}
L_{\rm
peak}=l\lambda_0M_{\bullet,0}\exp\bigg[\frac{l(1-\epsilon)\lambda_0}{\epsilon
c^2}t_{\rm peak}\bigg]\ ,
\end{equation}
where $M_{\bullet,0}$ is the seed BH mass at the triggering time
$t=0$. In eqn. (\ref{eqn:LC_model}), $\alpha$ determines how
rapidly the LC decays. Larger $\alpha$ values lead to more rapid
decay. Thus this model accommodates a broad range of possible
light curves. With this light curve model (\ref{eqn:LC_model}) we
have:
\begin{eqnarray}\label{eqn:BH_growth_M}
M_{\bullet}(L_{\rm peak},t) = & &
 \left\{
\begin{array}{lc}
\displaystyle \frac{L_{\rm peak}}{l\lambda_0}\exp\bigg[\frac{l(1-\epsilon)\lambda_0}{\epsilon c^2}(t-t_{\rm peak})\bigg], & 0\le t\le t_{\rm peak} \\ \\
\displaystyle\frac{L_{\rm
peak}}{l\lambda_0}+\frac{(1-\epsilon)L_{\rm peak}t_{\rm
peak}}{\epsilon
c^2}\times\\
\displaystyle\frac{1}{1-\alpha}\bigg[\bigg(\frac{t}{t_{\rm
peak}}\bigg)^{1-\alpha}-1\bigg], & t\ge t_{\rm peak}\ ,
\end{array} \right.
\end{eqnarray}
where we require $\alpha>1$ so that a BH cannot grow infinite
mass. The evolution of the luminosity Eddington ratio is
therefore:
\begin{eqnarray}\label{eqn:BH_growth_lambda}
\lambda(t) =
 \left\{
\begin{array}{lc}
\displaystyle \lambda_0, & 0\le t\le t_{\rm peak} \\ \\
\displaystyle\frac{(t/t_{\rm
peak})^{-\alpha}}{\displaystyle\frac{1}{\lambda_0}+\frac{l(1-\epsilon)t_{\rm
peak}}{\epsilon c^2(1-\alpha)}\bigg[\bigg(\frac{t}{t_{\rm
peak}}\bigg)^{1-\alpha}-1\bigg]}, & t\ge t_{\rm peak}\ ,
\end{array} \right.
\end{eqnarray}
where during the decaying part of the LC the Eddington ratio
monotonically decreases to zero. We assume that the seed BH mass
is a fraction $f$ of the peak mass $M_{\bullet}(t_{\rm
peak})\equiv L_{\rm peak}/(l\lambda_0)$, and we have
\begin{equation}\label{eqn:tpeak}
t_{\rm peak}=-\frac{\epsilon c^2}{l(1-\epsilon)\lambda_0}\ln
f=(-\ln f)t_{\rm Salpeter}\ .
\end{equation}
In what follows we set $f=10^{-3}$. Thus the seed BH is negligible
compared to the total mass accreted, and $t_{\rm peak}\approx
6.9t_{\rm Salpeter}$ where $t_{\rm Salpeter}$ is the ($e$-folding)
Salpeter timescale \citep{Salpeter_1964}. Although $\lambda_0=1$
is the formal definition of Eddington-limited accretion, it
assumes the electron scattering cross section, and in practice
super-Eddington accretion cannot be ruled out
\citep[e.g.,][]{Begelman_2002}. So we consider $\lambda_0$ within
the range $\log\lambda_0\in [-1,1]$.

The relic BH mass, given eqn. (\ref{eqn:BH_growth_M}), is simply
\begin{equation}\label{eqn:relic_mass1}
M_{\rm \bullet,relic} = \frac{L_{\rm
peak}}{l\lambda_0}+\frac{(1-\epsilon)L_{\rm peak}t_{\rm
peak}}{(\alpha-1)\epsilon c^2}=\frac{L_{\rm
peak}}{l\lambda_0}\bigg[1-\frac{\ln f}{\alpha-1}\bigg]\ .
\end{equation}
If instead we choose an exponential model for the decaying part of
the LC \citep[e.g.,][]{Kauffmann_Haehnelt_2000}, $L(t>t_{\rm
peak})=L_{\rm peak}\exp[-(t-t_{\rm peak})/t_{\rm QSO}]$, then the
relic mass becomes:
\begin{equation}
M_{\rm \bullet,relic} = \frac{L_{\rm
peak}}{l\lambda_0}+\frac{(1-\epsilon)L_{\rm peak}t_{\rm
QSO}}{\epsilon c^2}\ .
\end{equation}
Therefore LC models with large values of $\alpha$ mimic the
exponentially decaying model, and we do not consider the
exponential LC model further.


If the decaying parameter $\alpha$ is independent of mass, then
Eqns. (\ref{eqn:BH_growth_M}) and (\ref{eqn:relic_mass1}) imply
$M_{\rm \bullet,relic}\propto L_{\rm peak}$. In other words, the
scalings between $L_{\rm peak}$ and $M_{0}$, and between $M_{\rm
\bullet,relic}$ and $M_{0}$ should be the same for QSO-triggering
merger remnants (e.g., early type galaxies). Some theoretical
models and simulations predict $L_{\rm peak}\propto M_0^{4/3}$
\citep[e.g.,][]{King_2003,Di_Matteo_etal_2005,Springel_etal_2005b},
where momentum is conserved in self-regulated feedback; while
others invoking energy conservation predict $L_{\rm peak}\propto
M_0^{5/3}$ \citep[e.g.,][]{Silk_Rees_1998, Wyithe_Loeb_2003}. Two
recent determinations of the local black hole mass-halo mass
relation gave $M_{\rm \bullet,relic}\propto M_{0}^{1.65}$
\citep{Ferrarese_2002} and $M_{\rm \bullet,relic}\propto
M_{0}^{1.27}$ \citep{Baes_etal_2003}. Within uncertainties $M_{\rm
\bullet,relic}\propto L_{\rm peak}$ seems to be a reasonable
scaling, therefore we do not consider further mass dependence of
$\alpha$. Given the definition of $t_{\rm peak}$ (Eqn.\
\ref{eqn:tpeak}), our light curve model (\ref{eqn:LC_model}) is
then {\em universal} in the sense that it scales with $L_{\rm
peak}$ in a self-similar fashion.

\begin{figure}
  \centering
    \includegraphics[width=0.48\textwidth]{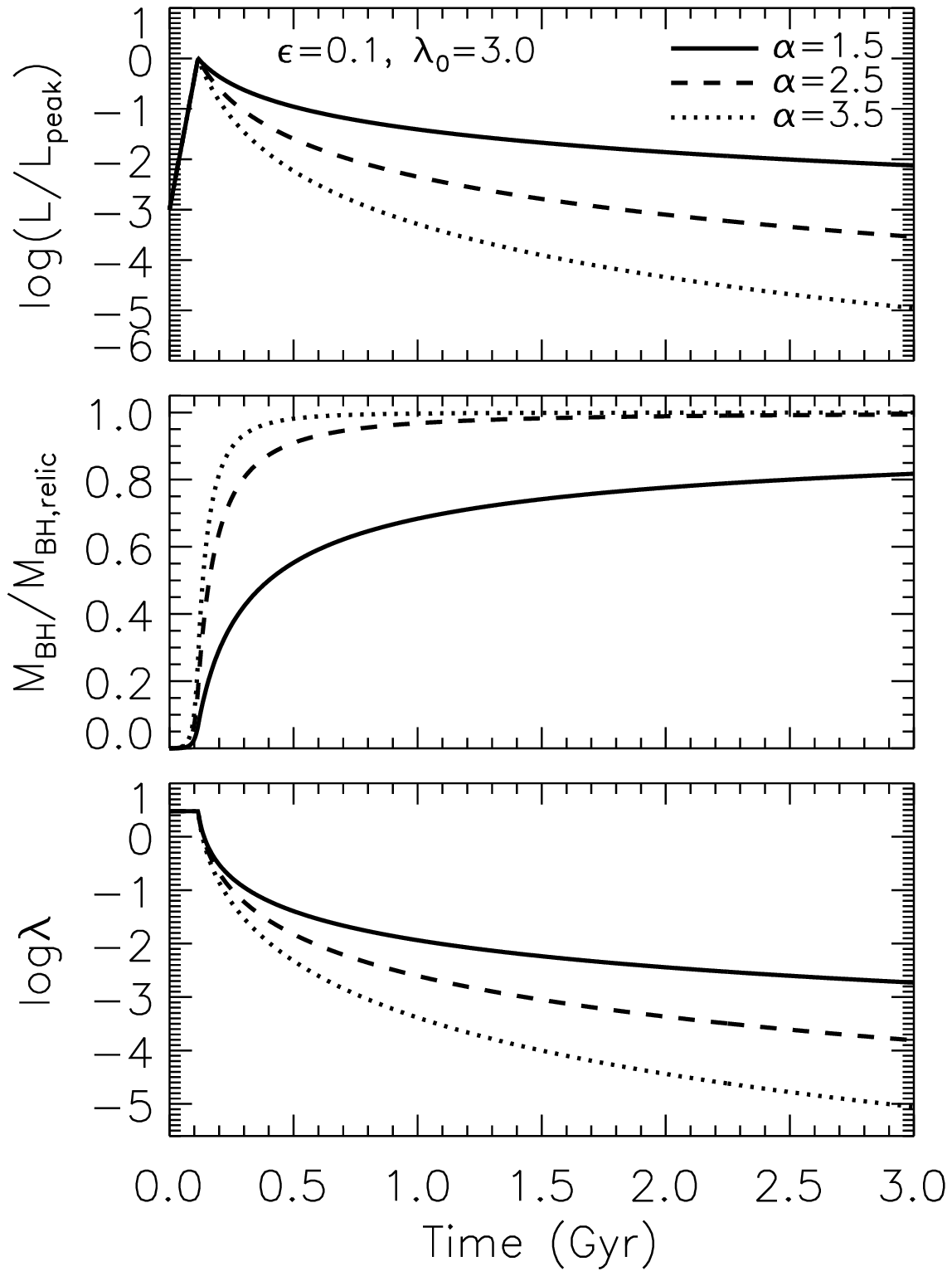}
    \caption{Example light curves for $\epsilon=0.1$ and
    $\lambda_0=3.0$, for $\alpha=1.5,2.5,3.5$. Larger values of $\alpha$ lead to more
    rapid decay. The time that the QSO spends above half of its peak luminosity is
    short, $<100\,$Myr for large values of $\alpha$.}
    \label{fig:LC_examples}
\end{figure}

As an example, Fig.\  \ref{fig:LC_examples} shows three light
curves with $\alpha=1.5,2.5,3.5$ for $\epsilon=0.1$ and
$\lambda_0=3.0$, in which case $t_{\rm peak}\sim 115\,{\rm Myr}$.
The time that a QSO spends above $50\%$ of its peak luminosity is
typically $\lesssim 100\,{\rm Myr}$ for sharp decaying curves, but
it can be substantially longer for extended decaying curves.

\subsection{BH Mass and Eddington Ratio
Distributions}\label{subsec:M_Edd_dist}

\begin{figure*}
  \centering
    \includegraphics[width=0.9\textwidth]{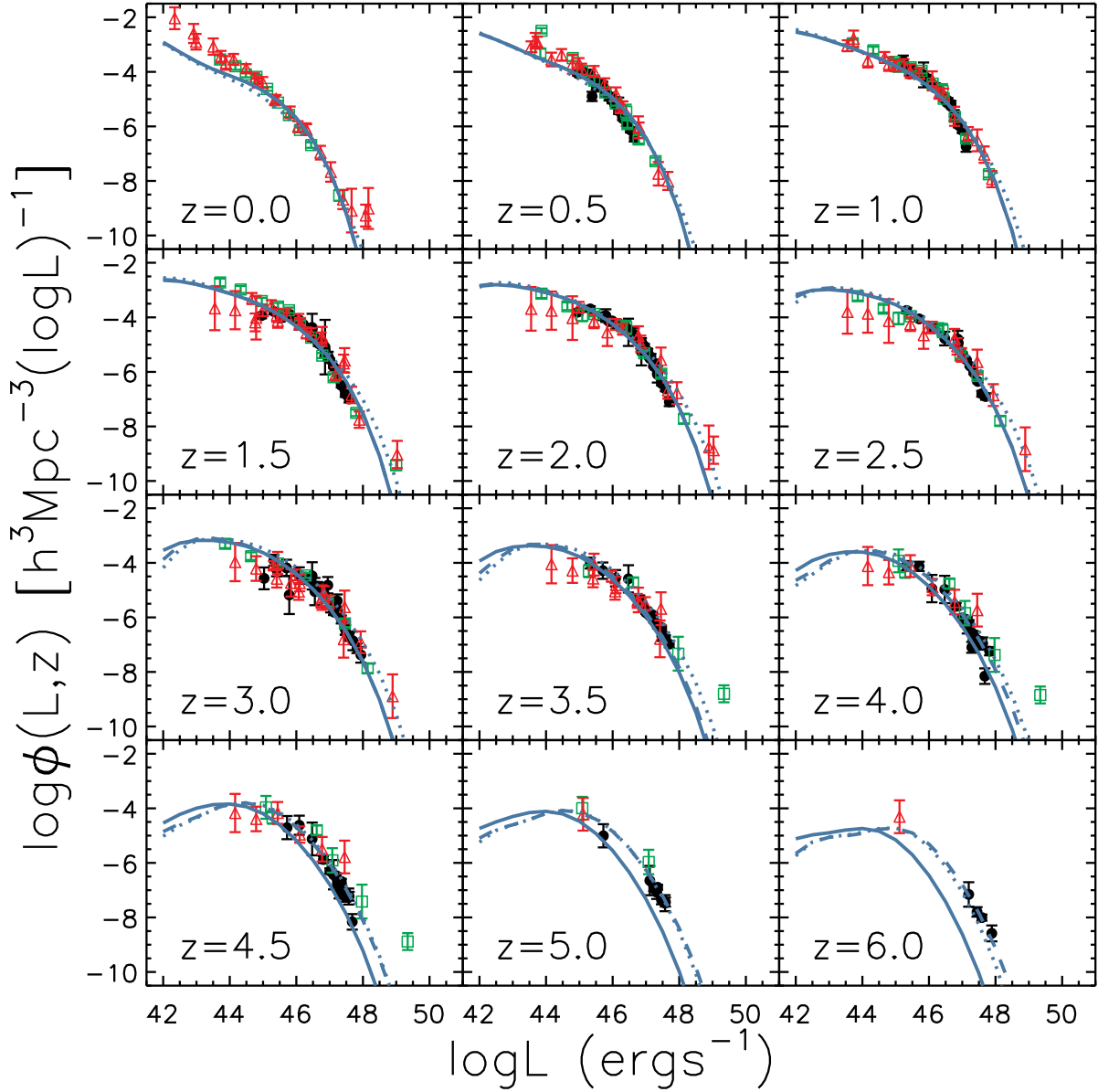}
    \caption{Predicted bolometric luminosity functions $\phi(L,z)\equiv d\Psi/d\log L$ in our
    fiducial model (solid lines). Overplotted are the complied bolometric LF data from \citet{Hopkins_etal_2007a} for
    optical (black circles), soft X-ray (green squares) and hard X-ray (red triangles) samples. The
    dashed lines are the predicted LF for a model where the normalization in the mean $L_{\rm peak}-M_0$
    relation is reduced by an additional amount at $z>3.5$, tuned to fit the LF at $z\ga 4.5$ (see \S\ref{subsec:disc1} for
    details). The blue dotted lines are the predictions based on an alternative prescription for the redshift evolution
    in the normalization of the $L_{\rm peak}-M_0$ relation as discussed in \S\ref{subsec:disc2}.}
    \label{fig:LFs}
\end{figure*}

Now we have specified the light curve model and the peak
luminosity-halo mass relation, and tied the luminosity function
$\Phi(L,z)$ to the QSO-triggering merger rate. We continue to
derive the instantaneous BH mass and Eddington ratio distributions
at fixed instantaneous luminosity $L$ and redshift $z$, which can
be compared with observationally determined distributions
\citep[e.g.,][]{Babic_etal_2007,Kollmeier_etal_2006,
Shen_etal_2008b,Gavignaud_etal_2008}.

Suppose that at redshift $z$ we observe quasars with instantaneous
luminosity $L$. These quasars consist of objects triggered at
different earlier redshift $z^\prime$ and are at different stages
of their evolution when witnessed at $z$ (e.g., Eqn.\
\ref{eqn:LF_merger}). There is a characteristic earlier redshift
$z_{c}$, determined by
\begin{equation}\label{eqn:zc}
t_{\rm age}(z) - t_{\rm age}(z_c) = t_{\rm peak}\ ,
\end{equation}
where $t_{\rm age}$ is the cosmic time. Quasars triggered between
$[z_c,z]$ are all in the rising part of the LC; while quasars
triggered earlier than $z_c$ are all in the decaying part of the
LC. In order to contribute to $\Phi(L,z)$, a quasar should have
peak luminosity $L_{\rm peak}=L$ if it is triggered right at
$z_c$, and it should have $L_{\rm peak}>L$ otherwise. Therefore
the triggering redshift distribution $dP(L_{\rm
peak},z^\prime|L,z)/dz^\prime$ (Eqn.\ \ref{eqn:prob_Lpeak_L})
peaks around\footnote{Except for low luminosities when the lower
halo mass cut $M_{\rm min}$ in the QSO-triggering rate starts to
kick in, the distribution $dP(L_{\rm
peak},z^\prime|L,z)/dz^\prime$ instead has a dip around $z_c$.}
$z^\prime\approx z_c$ since $B_{L}(L_{\rm peak},z^\prime)$
decreases rapidly when $L_{\rm peak}$ increases. Moreover, at the
observing redshift $z$, all the contributing quasars triggered
between $[z_c,z]$ will have Eddington ratio
$\lambda_{L,z^\prime\rightarrow z}=\lambda_0$ and BH mass
$M_{\bullet,L,z^\prime\rightarrow z}=L/(l\lambda_0)$; while all
the contributing quasars triggered before $z_c$ will have
Eddington ratio $\lambda_{L,z^\prime\rightarrow z}<\lambda_0$ and
BH mass $M_{\bullet,L,z^\prime\rightarrow z}>L/(l\lambda_0)$. The
probability distribution of instantaneous BH mass $M_{\bullet}$ at
the observing redshift $z$ and luminosity $L$ is therefore:
\begin{eqnarray}\label{eqn:BH_dist}
\frac{dP(M_{\bullet}|L,z)}{d\log M_{\bullet}}=\displaystyle
\frac{\displaystyle\frac{dP}{dz^\prime}\frac{dz^\prime}{d\log
M_{\bullet}}}{\displaystyle
1-\int_{z_c}^{z}\frac{dP}{dz^\prime}dz^\prime}\ ,\qquad
M_{\bullet}\ge \frac{L}{l\lambda_0}\ ,
\end{eqnarray}
where $dP(z^\prime|L,z)/dz^\prime$ is given by Eqn.\
(\ref{eqn:prob_Lpeak_L}) and $dz^\prime/dM_{\bullet}$ is
determined by {\em the decaying half} of the LC (Eqn.\
\ref{eqn:BH_growth_lambda}), i.e., the triggering redshift
$z^\prime (\ge z_c)$ is a monotonically increasing function of
$M_{\bullet}$ (note that here $M_{\bullet}$ refers to the
instantaneous BH mass observed at $z$). The denominator in the
above equation is to normalize the distribution -- we literally
augment the distribution with the pileup of objects with
$M_{\bullet}=L/(l\lambda_0)$ triggered within $[z_c,z]$. The
fraction of such objects is about $20\%$ for $L\gtrsim
10^{47}\,{\rm ergs^{-1}}$ and becomes negligible at lower
luminosities. This procedure removes the spike ($\delta$ function)
at the constant BH mass $M_{\bullet}=L/(l\lambda_0)$ in the
distribution, which is an artifact of our model\footnote{We have
tested with the alternative BH mass distribution at instantaneous
luminosity $L$ and redshift $z$ where the contribution of objects
with $M_{\bullet}=L/(l\lambda_0)$ is described by a $\delta$
function with normalization
$\int_{z_c}^{z}\frac{dP}{dz^\prime}dz^\prime$, and we find that
other derived distributions based on this BH mass distribution are
almost identical to those using Eqn.\ (\ref{eqn:BH_dist}).}. Eqn.\
(\ref{eqn:BH_dist}) gives the equivalent distribution of Eddington
ratios at $z$ and at instantaneous luminosity $L$.

Given the luminosity function (\ref{eqn:LF_merger}) and the
distribution of $M_{\bullet}$ (\ref{eqn:BH_dist}) we can determine
the {\em active} BH mass function (i.e., the mass function of active
QSOs above some luminosity cut $L_{\rm min}$) as
\begin{equation}\label{eqn:BH_mass_fun}
\frac{d\Psi_{M_\bullet}}{d\log M_{\bullet}}=\int_{L_{\rm
min}}^\infty \Phi(L,z)dL \frac{dP(M_{\bullet}|L,z)}{d\log
M_{\bullet}}\ .
\end{equation}


\section{The Reference Model}\label{sec:ref_model}

\begin{figure*}
  \centering
    \includegraphics[width=0.48\textwidth]{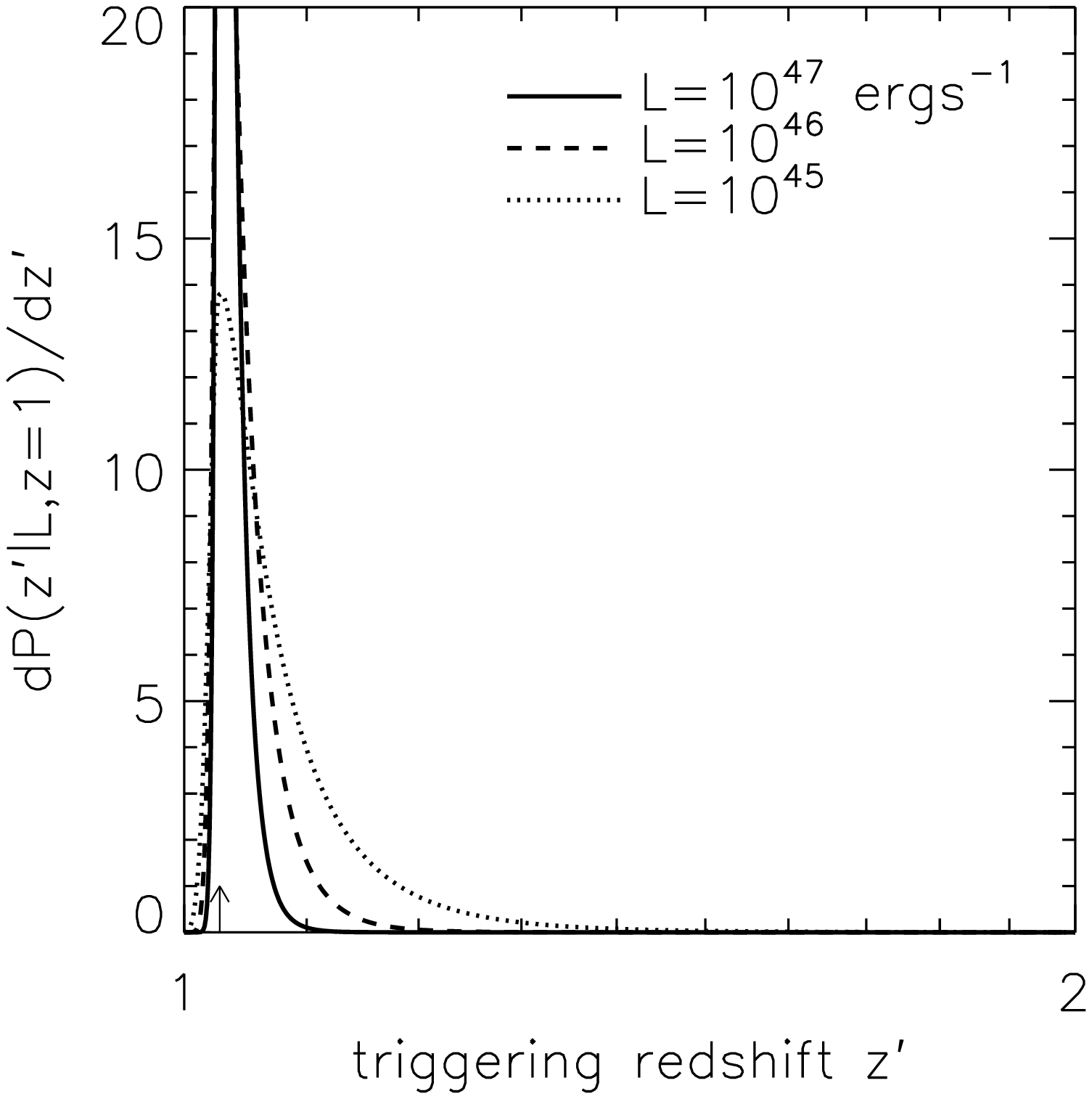}
    \includegraphics[width=0.48\textwidth]{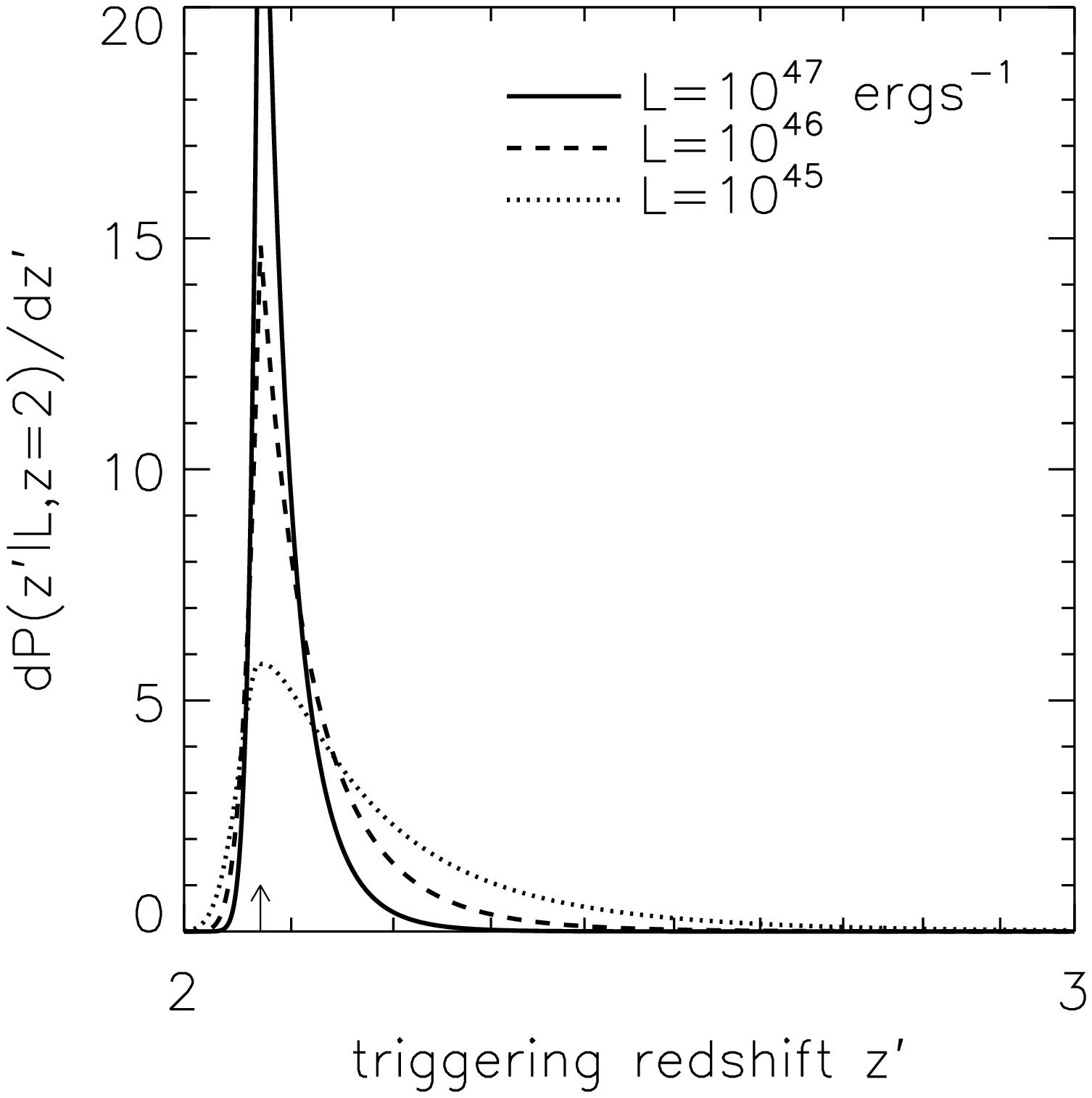}
    \includegraphics[width=0.48\textwidth]{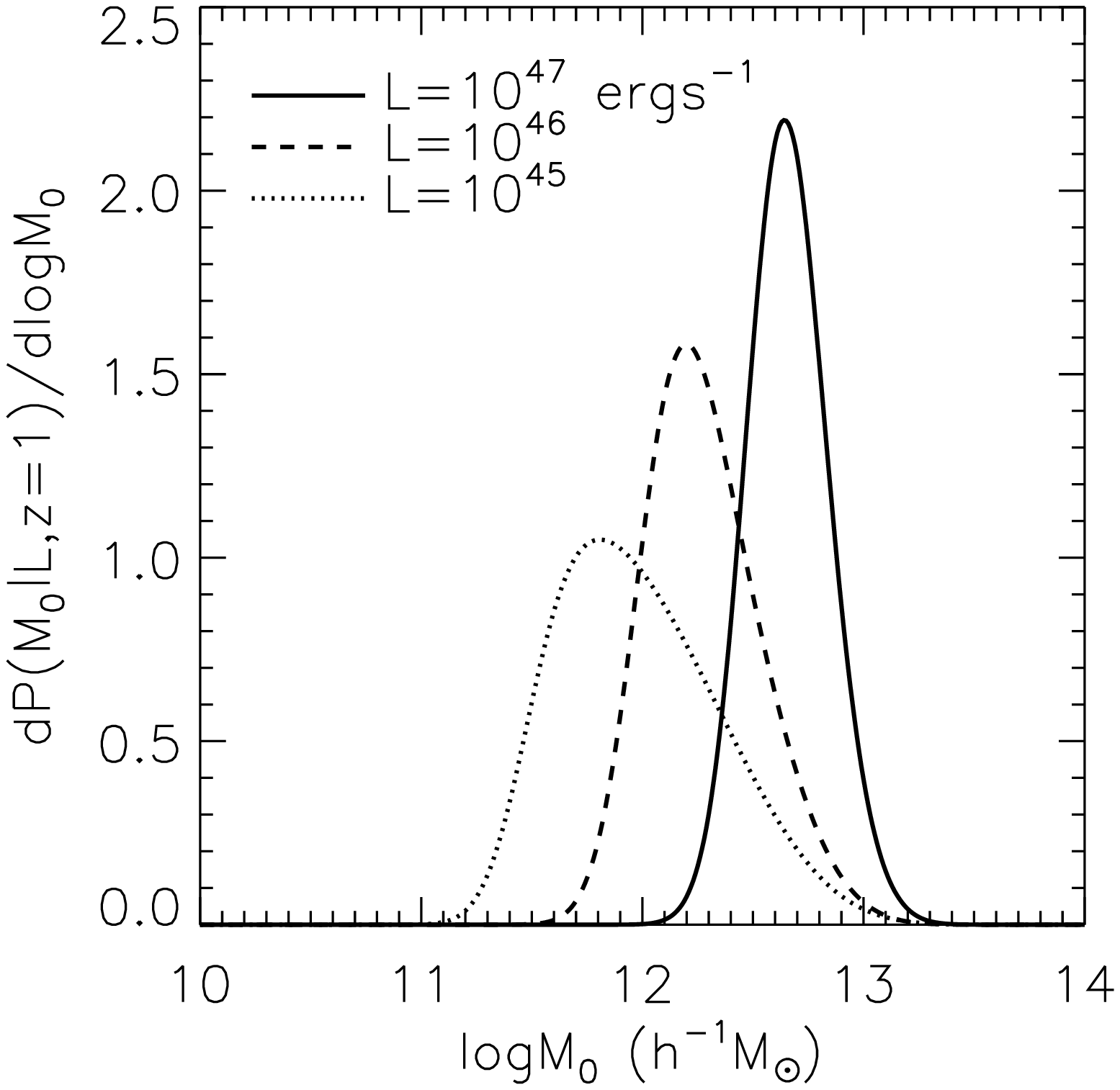}
    \includegraphics[width=0.48\textwidth]{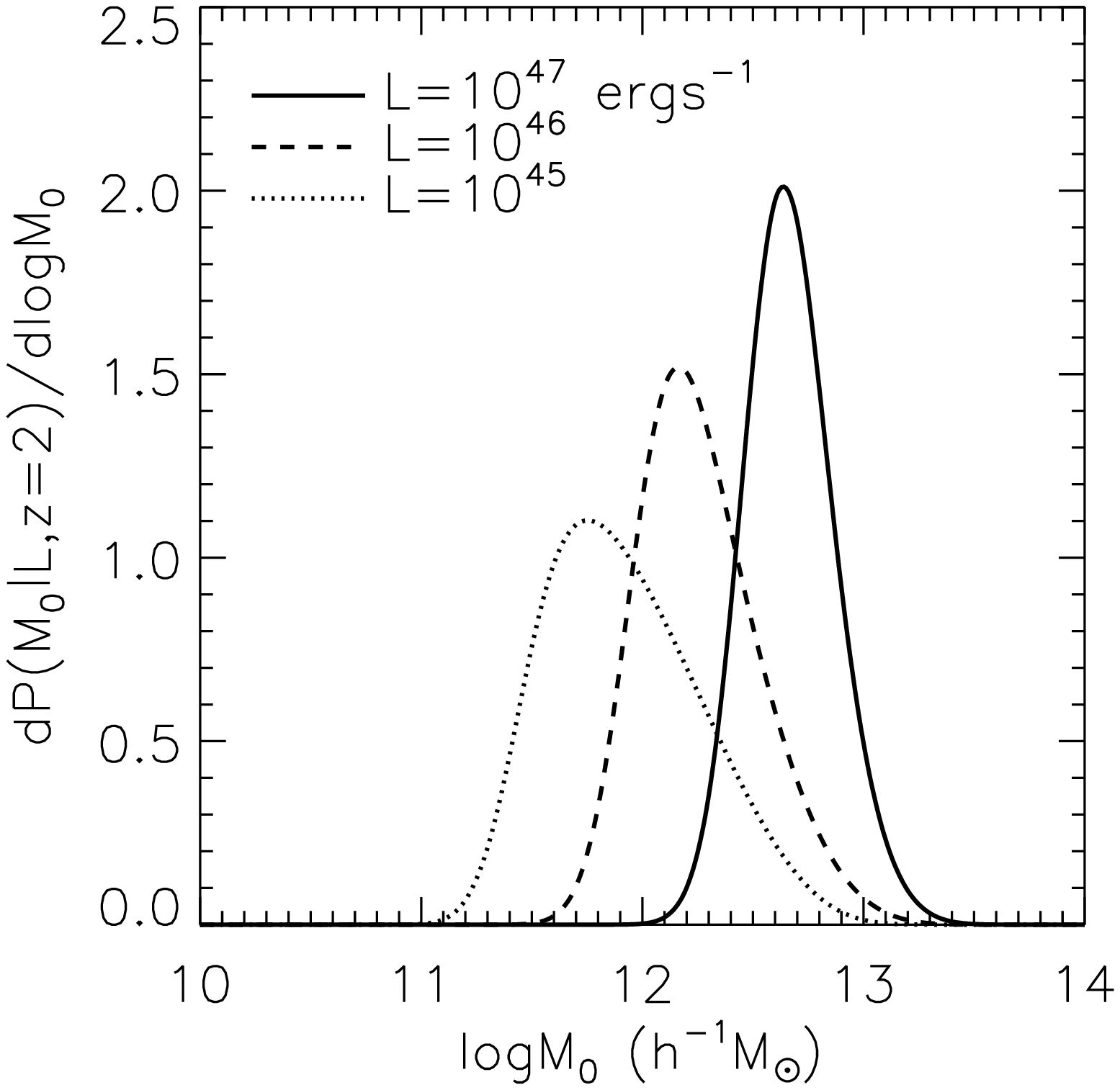}
    \caption{Distributions of triggering redshift (upper panels) and halo mass (lower panels) in our
    fiducial model, for three
    instantaneous luminosities and two values of observing redshifts. The distributions of triggering redshift generally peak around
    $z_c$ (marked by arrows) given by Eqn.\ (\ref{eqn:zc}). For lower luminosity, the distribution of the triggering
    redshift is more extended. As a consequence, there is a significant population of faded low-luminosity QSOs
    which have massive hosts. }
    \label{fig:zprime_dist}
\end{figure*}

We are now ready to convolve the LC model with the QSO-triggering rate Eqn.\
(\ref{eqn:agn_rate2}) to predict the LF $\Phi(L,z)$ using Eqn.\
(\ref{eqn:LF_merger}). Before we continue, let us review the model
parameters and available observational/theoretical constraints.

\begin{itemize}
\item {\em QSO triggering rate [$\xi_{\rm min}$, $M_{\rm min}(z)$,
$M_{\rm max}(z)$, ${\cal F}(z)$]:} We choose our fiducial value of
$\xi_{\rm min}\approx 0.3$ following the traditional definition of
major mergers ($1:3$ or $1:4$). For the minimum halo mass $M_{\rm
min}$ below which efficient accretion onto the BH is hampered we
simply choose $M_{\rm min}(z)=3\times 10^{11}\,h^{-1}M_\odot$. We
model the maximum halo mass cut above which the gas-rich merger
fraction drops as a function of redshift:
\begin{equation}
M_{\rm max}(z)=M_{\rm quench}(1+z)^\beta\ ,
\end{equation}
with $\beta> 0$. Therefore at lower redshift halos have a smaller
upper threshold. Since at $z<0.5$ rich group to cluster size halos
no longer host luminous QSOs we tentatively set $M_{\rm
quench}=1\times 10^{12}\,h^{-1}M_\odot$. The parameters $M_{\rm
min}$ and $M_{\rm max}$ control the the shape of the LF at both
the faint and bright luminosity ends. The global gas-rich merger
fraction, ${\cal F}(z)$, is more challenging to determine. Direct
observations of the cold gas fraction as function of redshift and
stellar mass (and therefore halo mass) are still limited by large
uncertainties. Moreover, how {\em gas-rich} a merger needs to be
in order to trigger efficient BH accretion is not well
constrained. For these reasons, we simply model ${\cal F}(z)$ as a
two-piece function such that ${\cal F}(z)=1$ when $z>2$, the peak
of the bright quasar population; and ${\cal F}(z)$ linearly
decreases to ${\cal F}_0$ at $z=0$. We note again that $f_{\rm
QSO}(z,M_0)$ should be regarded as the fraction of major mergers
that trigger QSO activity.




\item {\em The $L_{\rm peak}-M_0$ relation ($C$, $\gamma$,
$\sigma_L$)}: some merger event simulations
\citep[e.g.,][]{Springel_etal_2005b,Hopkins_etal_2005a,Lidz_etal_2006}
reveal a correlation between the peak luminosity and the mass of
the (postmerger) halo: $\langle L_{\rm peak}\rangle\approx 3\times
10^{45}(M_0/10^{12}\,h^{-1}M_\odot)^{4/3}\,{\rm ergs^{-1}}$ at
$z=2$, with a lognormal scatter $\sigma_L=0.35\,$dex. Other
analytical arguments predict $\gamma=5/3$, where energy is
conserved during BH feedback
\citep[e.g.,][]{Silk_Rees_1998,Wyithe_Loeb_2003}. We choose
$\gamma$ values between $[4/3,5/3]$ since the above simulations
are also consistent with the $\gamma=5/3$ slope. Furthermore, we
allow the normalization parameter $C$ to evolve with redshift,
$C(z)=C(z=0)+\log[(1+z)^{\beta_1}]$ with $\beta_1>0$. Thus for the
same peak luminosity, the host halos become less massive at higher
redshift. This decrease of the characteristic halo mass with
redshift is also modeled by several other authors, although we do
not restrict to $\beta_1= 1$ \citep[e.g.,][]{Lapi_etal_2006}, or
more complicated prescriptions
\citep[e.g.,][]{Wyithe_Loeb_2003,Croton_2009}, since the form of
the $L-M_0$ (or $M_\bullet-M_0$) relation is also different in
various prescriptions. A smaller characteristic halo mass is
easier to account for the QSO abundance at high redshift since
there are more mergers of smaller halos, but it also reduces the
clustering strength. The intrinsic scatter of the $L_{\rm
peak}-M_0$ relation, $\sigma_L$, will have effects on both the
luminosity function and clustering. Larger values of $\sigma_L$
lead to higher QSO counts, but will also dilute the clustering
strength due to the up-scattering of lower mass halos
\citep[e.g.,][]{White_etal_2008}. The $L_{\rm peak}-M_0$ relation
establishes a baseline for the mapping from halos to SMBHs.
Although in our fiducial model we adopt the above rather simple
scaling [$\propto (1+z)^{\beta_1}$] for the redshift evolution in
$C(z)$, we will discuss alternative parameterizations for $C(z)$
in \S\ref{subsec:disc1} and \S\ref{subsec:disc2}.

\item {\em The light curve model ($\lambda_0$, $\epsilon$, $t_{\rm
peak}$, $\alpha$):} our chosen fiducial value for $t_{\rm peak}$
is $t_{\rm peak}=6.9t_{\rm Salpeter}$ ($f=10^{-3}$); but our
results are insensitive to the exact value of $f$. The radiative
efficiency is $\epsilon\approx 0.1$ from the Soltan argument
(Eqn.\ \ref{eqn:soltan}). We choose $\lambda_0$ between $[0.1,10]$
and $\alpha>1.1$.



\end{itemize}



Normally to find the best model parameters one needs to perform a
$\chi^2$ minimization between model predictions and observations.
We do not perform such exercise here because it is difficult to
assign relative weights to different sets of observations (i.e.,
LF, quasar clustering, Eddington ratio distributions, etc), and we
don't know well enough the systematics involved. Instead, we
experiment with varying the model parameters within reasonable
ranges, to achieve a global ``good'' (as judged by eye) fit to the
overall observations. Our fiducial model has the following
parameter values: $\xi_{\rm min}=0.25$, ${\cal F}_0=1.0$,
$\beta=1.5$, $\gamma=5/3$, $C(z=0)=\log(6\times
10^{45})-12\gamma$, $\beta_1=0.2\gamma$, $\sigma_L=0.28$,
$\lambda_0=3.0$ and $\alpha=2.5$, which are all within reasonable
ranges. In particular we found that the parameterization of ${\cal
F}(z)$ is unnecessary because the effect of cold gas consumption
is somewhat described by $M_{\rm max}(z)$ already. Below we
describe in detail the predictions of this reference model and
comparison with observations, and we defer the model variants and
caveats to \S\ref{sec:disc}.

Fig.\  \ref{fig:LFs} shows the model LF $d\Psi/d\log L\equiv
L\ln(10)\Phi(L,z)$ in solid lines, given by Eqn.\
(\ref{eqn:LF_merger}), where we overplot the compiled bolometric
luminosity function data in \citet[][]{Hopkins_etal_2007a}. In
computing Eqn.\ (\ref{eqn:LF_merger}) we integrate up to $z=20$
but the integral converges well before that. The model
under-predicts the counts at the faint luminosity end
($L<10^{45}\,{\rm ergs^{-1}}$) for $z<0.5$, leaving room for low
luminosity AGNs triggered by mechanisms other than a major merger
event (i.e., by secular processes). At redshift $z\gtrsim 4.5$,
the model also underestimates the QSO abundance. This could be
alleviated if the characteristic halo mass shifts to even lower
values, i.e., even larger values of the parameter $C$ at $z\gtrsim
4.5$ (see discussions in \S\ref{subsec:disc1} and
\S\ref{subsec:disc2}). Alternatively, the high-$z$ SMBH population
may be different in the sense that it is not tied to $\xi\gtrsim
0.3$ major mergers events directly. Also, the halo merger rate
(Eqn.\ \ref{eqn:halo_merger_rate}) has been extrapolated to such
high redshifts for the massive halos considered here. Despite
these facts, the model correctly reproduces the LF from $z\approx
0.5$ to $z\approx 4.5$. The turnover below $L\approx 10^{44}\,{\rm
ergs^{-1}}$ at $z\gtrsim 2.5$ is caused by the lower mass cutoff
$M_{\rm min}=3\times 10^{11}\,h^{-1}M_\odot$. At lower redshift,
this turnover flattens due to the gradual pile up of evolved
high-peak luminosity quasars well after $t_{\rm peak}$.

We show some of the predicted distributions in Fig.\
\ref{fig:zprime_dist} for this reference model. The upper panels
show the distributions of the triggering redshift $z^\prime$ for
instantaneous luminosities $L=10^{45},10^{46},10^{47}\,{\rm
ergs^{-1}}$ at $z=1$ (upper left) and $z=2$ (upper right). As
expected, the distribution of the triggering redshift peaks around
the characteristic redshift $z_c$ given by Eqn.\ (\ref{eqn:zc}).
The bottom panels show the distributions of host halo mass $M_0$
for the three instantaneous luminosities, at the two redshifts
respectively. For bright quasars ($L>10^{46}\,{\rm ergs^{-1}}$),
the distributions of $M_0$ are roughly log-normal, with the width
and mean determined mainly through the $L_{\rm peak}-M_0$ relation
-- but also slightly modified through the convolutions with other
distributions (see Eqns.\ \ref{eqn:prob_Lpeak_L},
\ref{eqn:prob_M_Lpeak} and \ref{eqn:prob_M_L}). For faint QSOs
($L\la 10^{45}\,{\rm ergs^{-1}}$), however, the halo mass
distribution has a broad high-mass tail contributed by evolved
high-peak luminosity quasars triggered earlier. This contamination
of massive halos hosting low luminosity QSOs will increase the
clustering bias of the faint quasar population.

\begin{figure}
  \centering
    \includegraphics[width=0.48\textwidth]{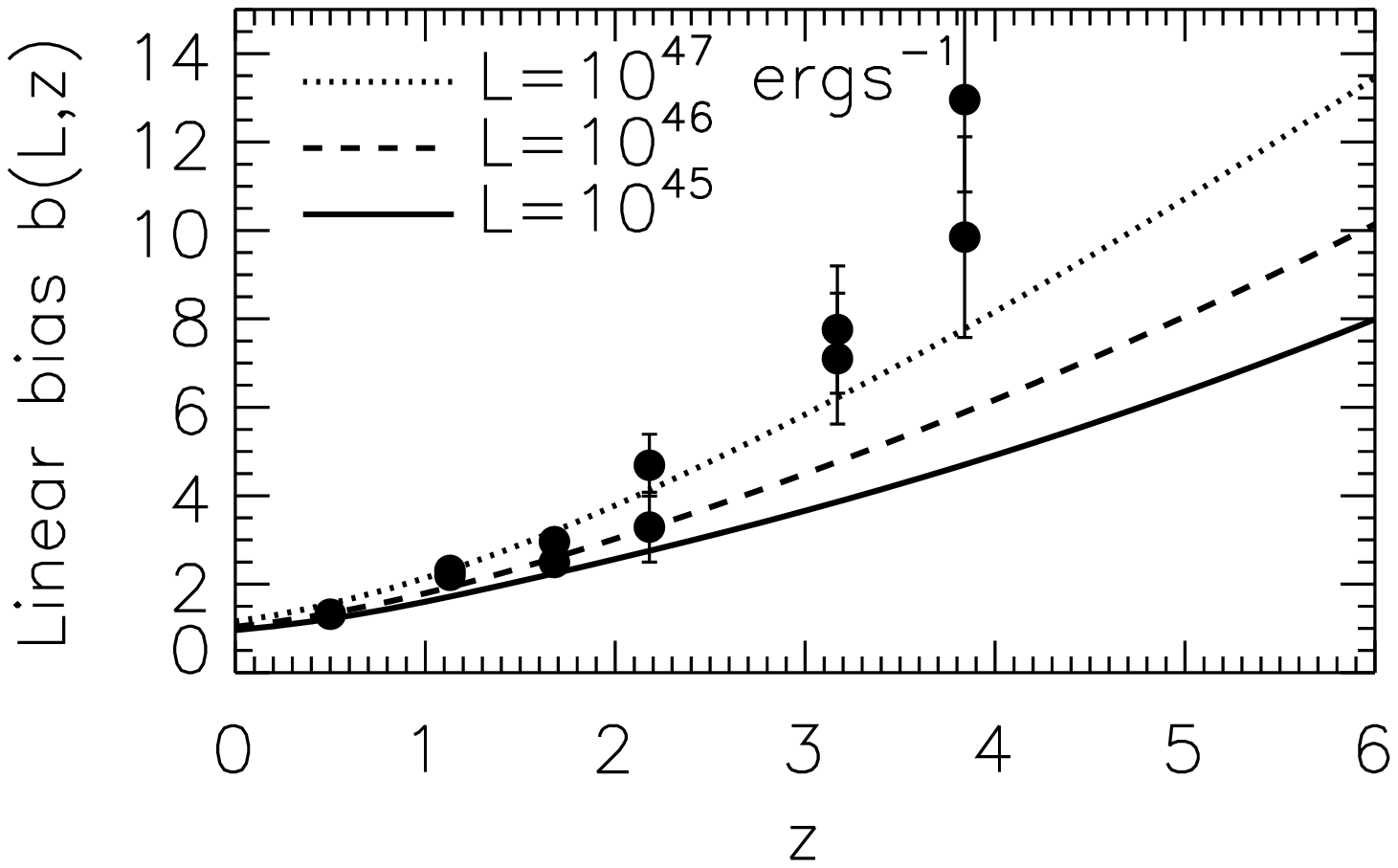}
    \includegraphics[width=0.48\textwidth]{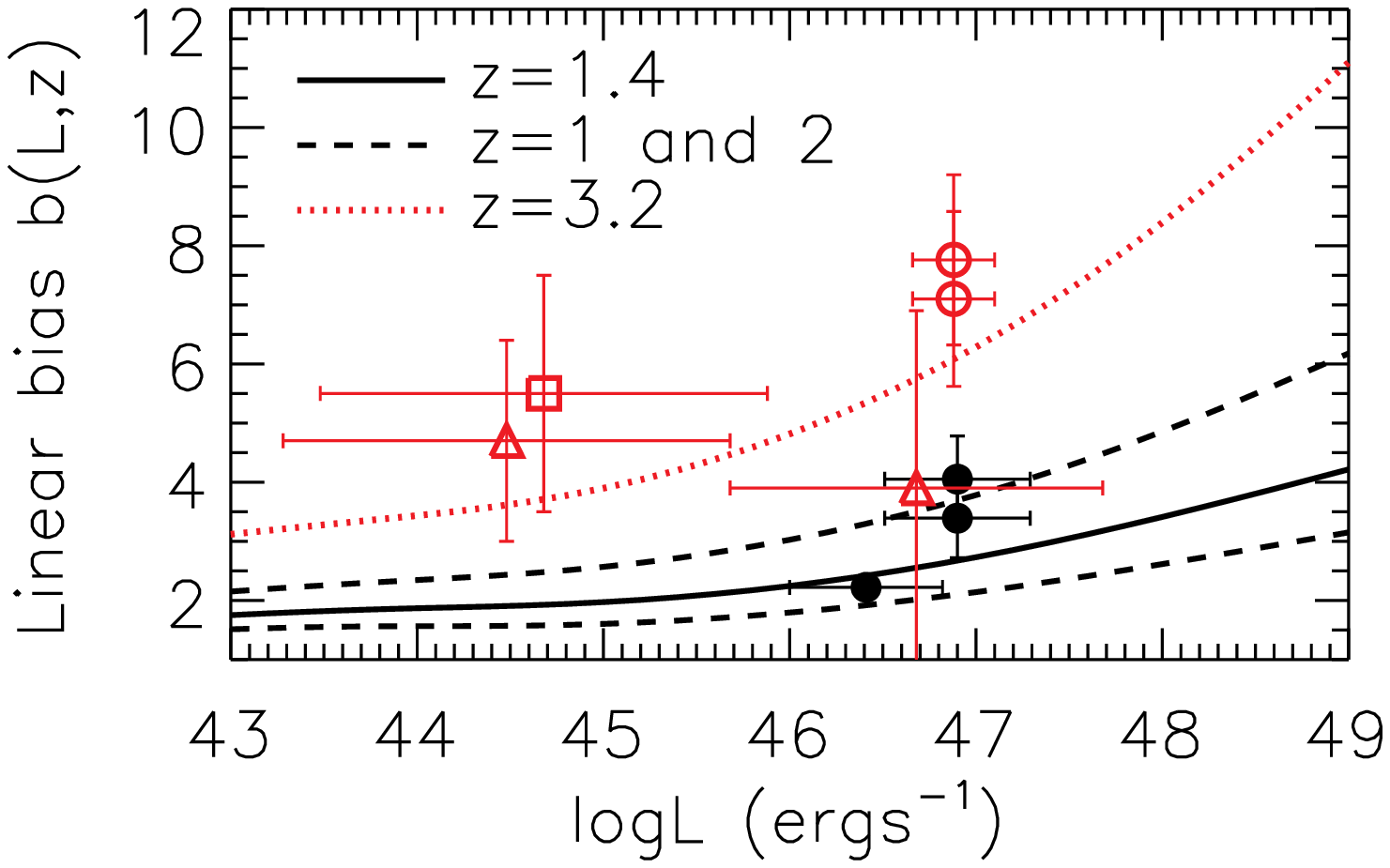}
    \caption{Predictions for the linear bias of quasar clustering in our fiducial model. {\em Upper}:
    bias evolution with redshift. The data points are from the measurements in \citet{Shen_etal_2009a} and three lines show
    the predicted bias for quasars with instantaneous luminosity $L=10^{45},10^{46},10^{47}\,{\rm ergs^{-1}}$. The median
    luminosity of quasars used in the clustering analysis for the six redshift bins are: $\langle\log(L/{\rm ergs^{-1}})\rangle=45.6,
    46.3,46.6,46.9,46.9,47.1$. {\em Bottom:} predicted luminosity dependence of quasar bias. The filled circles are
    from the measurements in \citet{Shen_etal_2009a} for the $10\%$ most luminous and the remainder of a sample of
    quasars spanning $0.4<z<2.5$. The open circles are the measurement of quasar clustering for $2.9<z<3.5$ in
    \citet{Shen_etal_2009a}. The open square is from \citet{Francke_etal_2008} using cross-correlation of X-ray selected
    AGN with high redshift galaxies, and the open triangles are from \citet{Adelberger_Steidel_2005b} using cross-correlation
    of optical AGNs with galaxies. Note that for the quasar clustering data we plot biases derived from both with and
    without including negative correlation function data in the fitting \citep[see table 1 of ][]{Shen_etal_2009a}.}
    \label{fig:bias}
\end{figure}

\begin{figure*}
  \centering
    \includegraphics[width=0.33\textwidth]{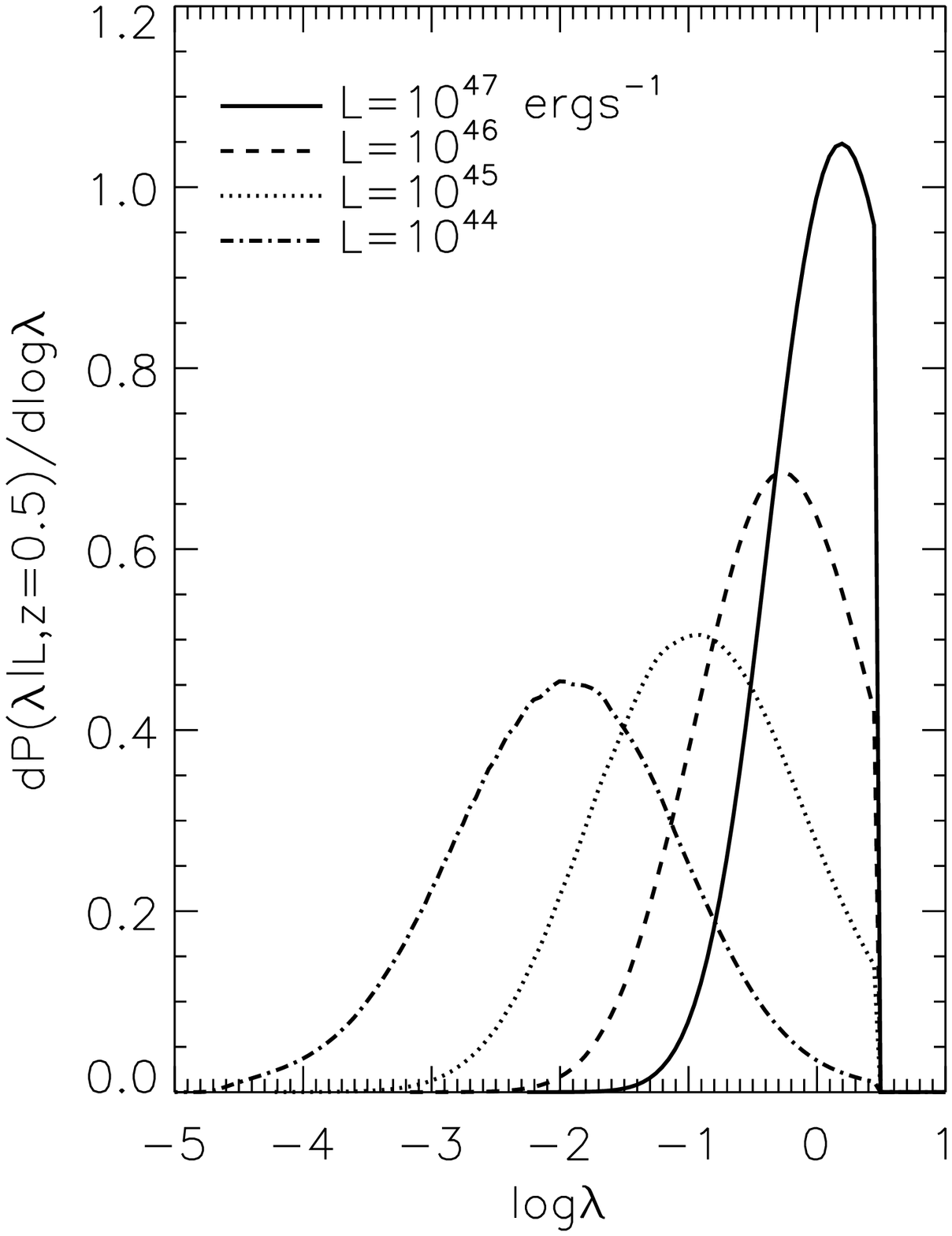}
    \includegraphics[width=0.33\textwidth]{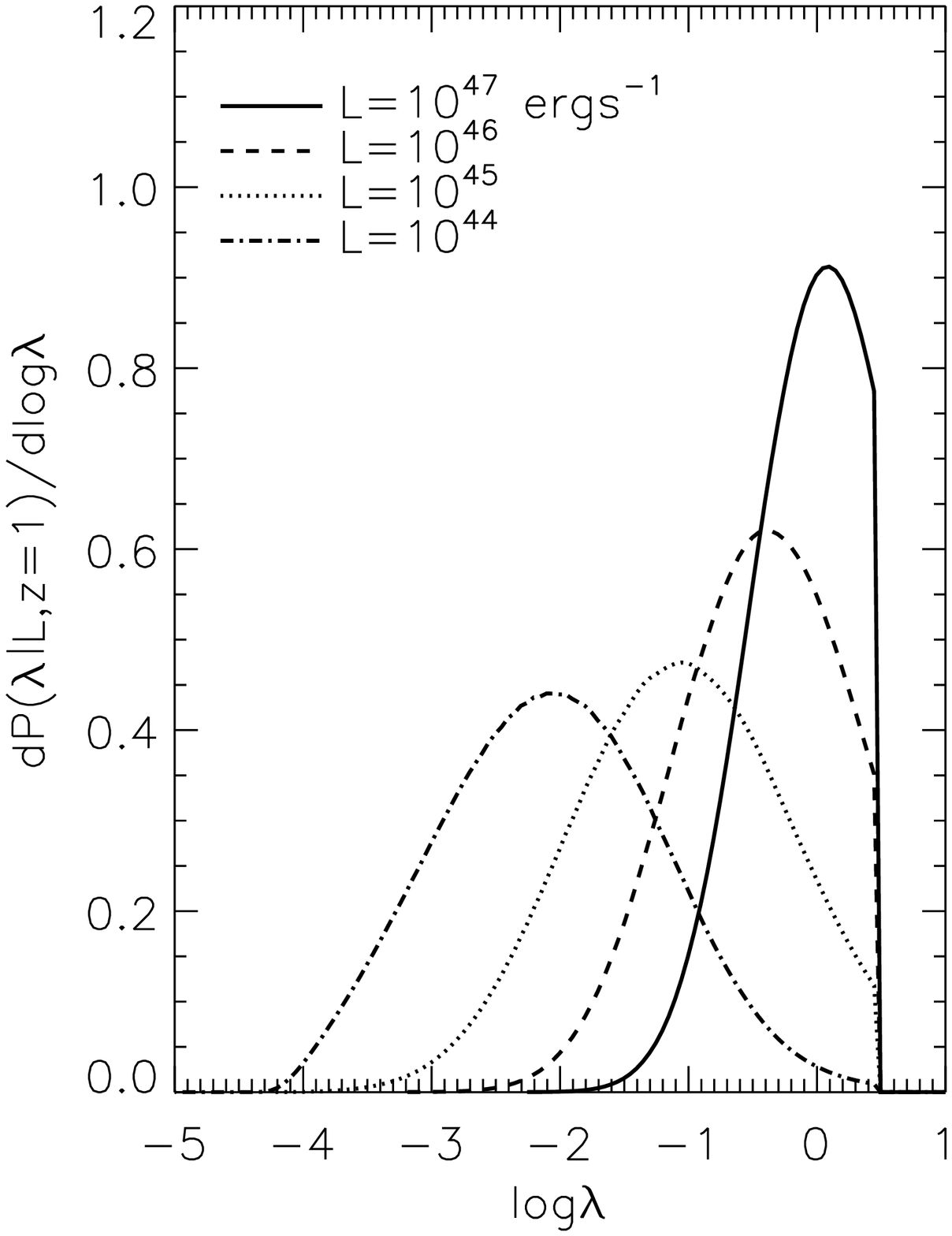}
    \includegraphics[width=0.33\textwidth]{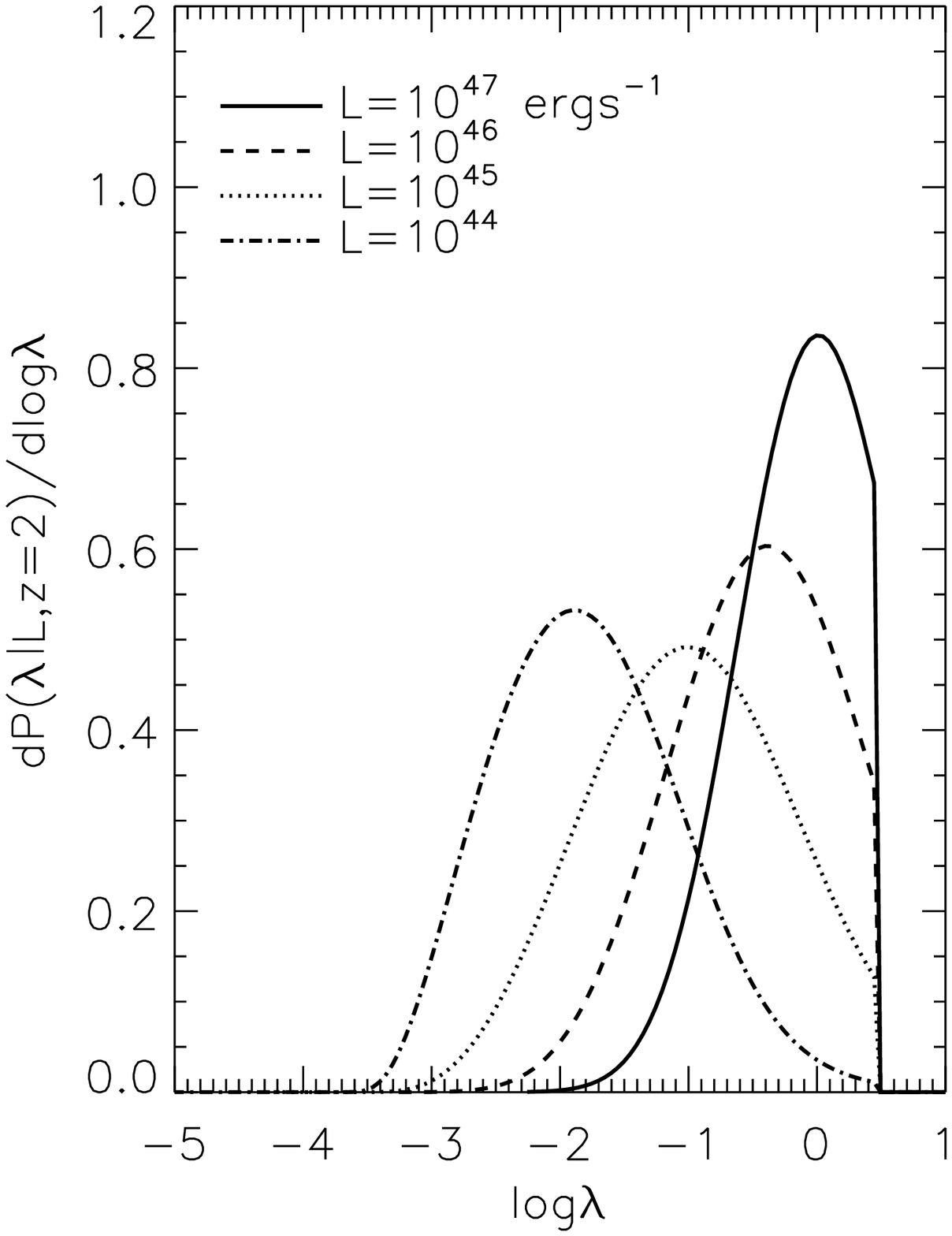}
    \caption{Predictions for the Eddington ratio distributions in our fiducial model for three different redshifts.
     The distribution is narrower and has higher peak values towards higher luminosities. These distributions
     are in good agreement with the observed distributions for both bright quasars and low luminosity
     AGNs \citep[e.g.,][]{Kollmeier_etal_2006,Babic_etal_2007,Shen_etal_2008b,Gavignaud_etal_2008}.}
    \label{fig:Edding_dist}
\end{figure*}

Fig.\  \ref{fig:bias} shows our model predictions for the redshift
and luminosity dependence of quasar bias, compared with
observations. The model-predicted redshift evolution of quasar
bias is in good agreement with observations
\citep[e.g.,][]{PMN_2004,Croom_etal_2005,Porciani_Norberg_2006,
Myers_etal_2006,Myers_etal_2007a,Myers_etal_2007b,Shen_etal_2007b,
da_Angela_2008,Padmanabhan_etal_2008,Shen_etal_2008a,Shen_etal_2009a,Ross_etal_2009}.
But it has some difficulties in accounting for the large bias at
$z\approx 4$, resulting from the need to reproduce the quasar
abundance at such high redshift (i.e., the characteristic halo
mass shifts to lower values). We discuss possible solutions to
this problem in \S\ref{subsec:disc1}. Our model also predicts the
luminosity dependence of quasar clustering. At low luminosities,
quasar bias depends weakly on luminosity, which is consistent with
observations
\citep[e.g.,][]{Porciani_Norberg_2006,Myers_etal_2007a,da_Angela_2008,Shen_etal_2009a}.
This simply reflects the fact that quasars are not light bulbs,
i.e., there is non-negligible scatter around the mean $L_{\rm
peak}-M_0$ relation, and some faint quasars live in massive halos
because of the evolving light curve
\citep[e.g.,][]{Adelberger_Steidel_2005b,Hopkins_etal_2005a,Lidz_etal_2006}.
On the other hand, the model predicts that the bias increases
rapidly towards high luminosities and/or at high redshift. This is
consistent with the findings by \citet{Shen_etal_2009a} that the
most luminous quasars cluster more strongly than intermediate
luminosity quasars at $z<3$. Our model-predicted luminosity
dependent quasar clustering broadly agrees with the measurements
in \citet[][]{Shen_etal_2009a} for $0.4<z<2.5$, but we caution
that their measurements are done for samples with a broad redshift
and luminosity range in order to build up statistics (e.g., see
their fig. 2), hence a direct comparison is somewhat difficult. At
redshift $\sim 3$, \citet{Adelberger_Steidel_2005b} and
\citet[][]{Francke_etal_2008} measured the clustering of low
luminosity AGN using cross-correlation with galaxies. Their data
are shown in Fig.\ \ref{fig:bias} in open square and triangles for
the measurements in \citet{Francke_etal_2008} and
\citet{Adelberger_Steidel_2005b} respectively, which are broadly
consistent with our predictions. However the clustering of
optically bright quasars is apparently at odds with the low bias
value derived from the AGN-galaxy cross-correlation in
\citet[][]{Adelberger_Steidel_2005b}, likely caused by the fact
that the latter AGN sample spans a wide redshift range $1.6\la
z\la 3.7$ and luminosity range, and that the uncertainty in their
bias determination is large.

Fig.\  \ref{fig:Edding_dist} shows our model predictions for the
Eddington ratio distributions at various redshifts and
instantaneous luminosities. Aside from the cutoff at $\lambda_0=3$
(our model setup), these Eddington ratio distributions are
approximately log-normal, and broaden towards fainter
luminosities. This again reflects the nature of the light curve --
at lower luminosities, there are more objects at their late
evolutionary stages and shining at lower Eddington ratios. These
predictions are in good agreement with the observed Eddington
ratio distributions for bright quasars ($L\gtrsim 10^{46}\,{\rm
ergs^{-1}}$) where the distribution is narrow and peaks at high
mean values ($\langle\log\lambda\rangle\in [-1,0]$)
\citep[e.g.,][]{Woo_Urry_2002,Kollmeier_etal_2006,Shen_etal_2008b},
as well as for faint AGNs ($L\lesssim 10^{45}\,{\rm ergs^{-1}}$)
where the distribution is broader and peaks at lower mean values
($\langle\log\lambda\rangle\in [-3,-1]$)
\citep[e.g.,][]{Babic_etal_2007,Gavignaud_etal_2008}. However the
predicted mean values and widths are not necessarily exactly the
same as those determined from observations, since uncertainties in
the BH mass estimators used in these observations may introduce
additional scatter and biases in the observed Eddington ratio
distributions \citep[e.g.,][]{Shen_etal_2008b}. The Eddington
ratio distributions of faint AGNs ($L\lesssim 10^{44}\,{\rm
ergs^{-1}}$) are particularly broad at low redshift since more
high-$L_{\rm peak}$ objects have had enough time to evolve. On the
other hand, at high redshift $z\gtrsim 2$, the minimum allowable
Eddington ratio is set by the cosmic age at that redshift, e.g.,
$\lambda_{\rm min}\approx 10^{-4}$ at $z=2$ using Eqn.\
(\ref{eqn:BH_growth_lambda}). This explains the narrowing of the
lower end of the Eddington ratio distribution for $L=10^{44}\,{\rm
ergs^{-1}}$, seen in the right panel of Fig.\
\ref{fig:Edding_dist}.

\begin{figure}
  \centering
    \includegraphics[width=0.48\textwidth]{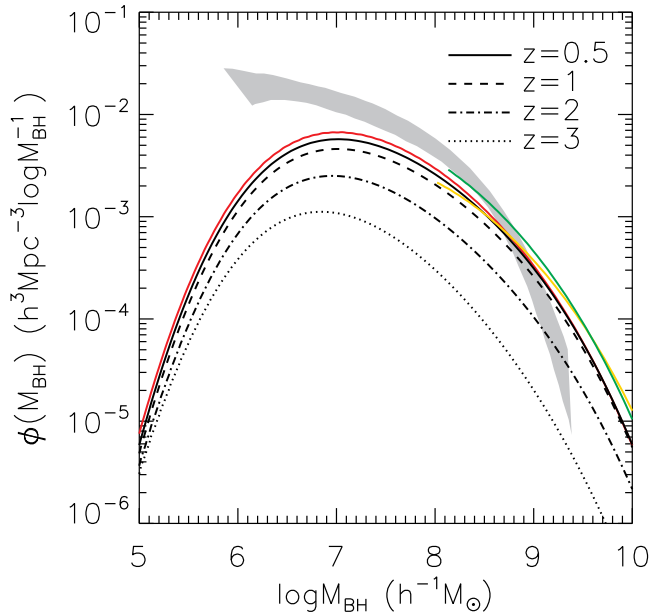}
    \caption{BHMFs assembled at various redshifts in our fiducial model. The gray shaded region shows
    the estimates for the local BHMF from \citet{Shankar_etal_2009a}.
    The red line shows the prediction for the local BHMF, which is
    incomplete at $M_{\bullet}\lesssim 10^{7.5}\, M_\odot$ by a factor of a few because we did not include contributions
    from AGNs triggered by secular processes or minor mergers (as reflected in the failure to reproduce the LF at the
    low luminosity end $L<10^{45}\,{\rm ergs^{-1}}$ at $z<0.5$, see Fig.\  \ref{fig:LFs}). The yellow and
    green lines show the predicted local BHMF at $M_\bullet>10^8\, h^{-1}M_\odot$ after correcting for
    BH coalescence; these corrections are likely upper limits (see \S\ref{subsec:BH_Coalescence} for details).}
    \label{fig:act_BMF}
\end{figure}

Since we have reproduced the observed luminosity function and
predicted the Eddington ratio distributions as function of
luminosity, we can use Eqn.\ (\ref{eqn:BH_mass_fun}) to derive the
{\em active} BHMF in QSOs above some minimum luminosity. In Fig.\
\ref{fig:act_BMF} we show the {\em total} BHMF (setting $L_{\rm
min}=0$ in Eqn.\ \ref{eqn:BH_mass_fun}) assembled at several
redshifts, where the gray shaded region indicates the estimates of
the local dormant BHMF based on various galaxy bulge-BH scaling
relations \cite[][]{Shankar_etal_2009a}. Our model prediction for
the local BHMF is incomplete at the low-mass end, mainly because
our model does not include low luminosity AGN activity (presumably
linked to $M_{\bullet}\lesssim 10^7\,M_\odot$ BH growth) possibly
triggered by secular processes (see further discussion in
\S\ref{subsec:disc1}), and also because we set the minimal halo
mass that can trigger QSO activity during major mergers $M_{\rm
min}=3\times10^{11}\,h^{-1}M_\odot$ in order to reproduce the
faint-end LF at high redshift. At the high mass end, the predicted
slope in the {\em total} BHMF broadly agrees with (although is
shallower than) that for the local dormant BHMF.
The majority of the present-day $\gtrsim 10^{8.5}\,M_\odot$ BHs
were already in place by $z=1$, but less than $50\%$ of them were
assembled by $z=2$. This is somewhat in disagreement with
\citet[][]{McLure_Dunlop_2004}, who claimed that the majority of
$>10^{8.5}\,M_\odot$ SMBHs are already in place at $z\sim 2$ based
on virial BH mass estimates of optically selected bright quasars.
We suspect this discrepancy is caused by: 1) the fact that virial
BH mass estimates tend to systematically overestimate the true BH
masses due to a Malmquist-type bias \citep[see discussions in
][]{Shen_etal_2008b}; 2) the high-mass end slope in the local
dormant BHMF is steeper than our model predictions. On the other
hand, our model predictions are in good agreement with the
predictions by \citet[][]{Shankar_etal_2009a} for the high-mass
end. It is possible to make our model prediction agree with the
local BHMF at the high-mass end by imposing some cutoff in the
light curve [Eqn.\ (\ref{eqn:LC_model})] such that BHs cannot grow
too massive; but a more accurate observational determination of
the high-mass end of the local BHMF is needed to resolve these
issues.

\begin{figure*}
  \centering
    \includegraphics[width=0.48\textwidth]{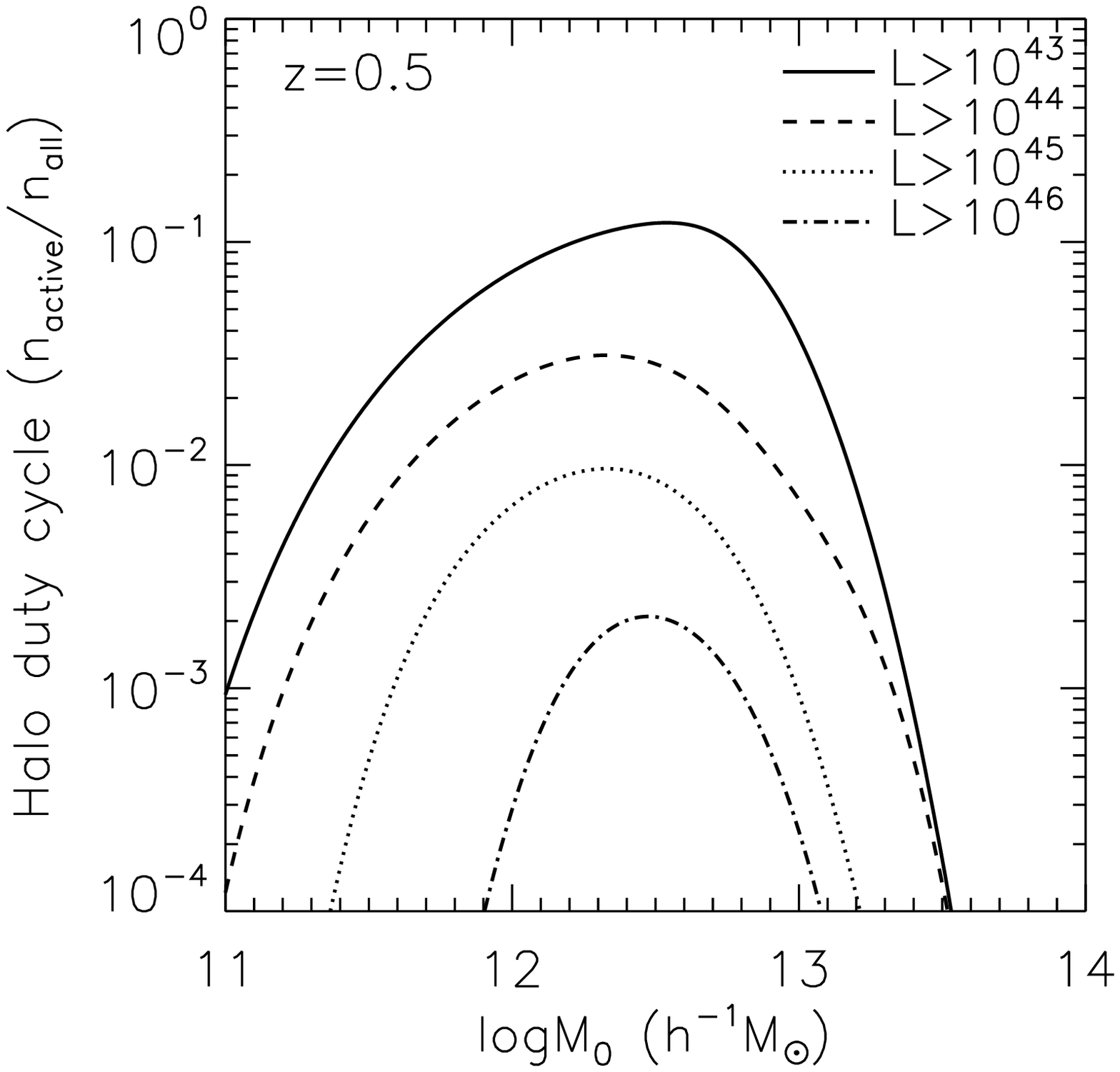}
    \includegraphics[width=0.48\textwidth]{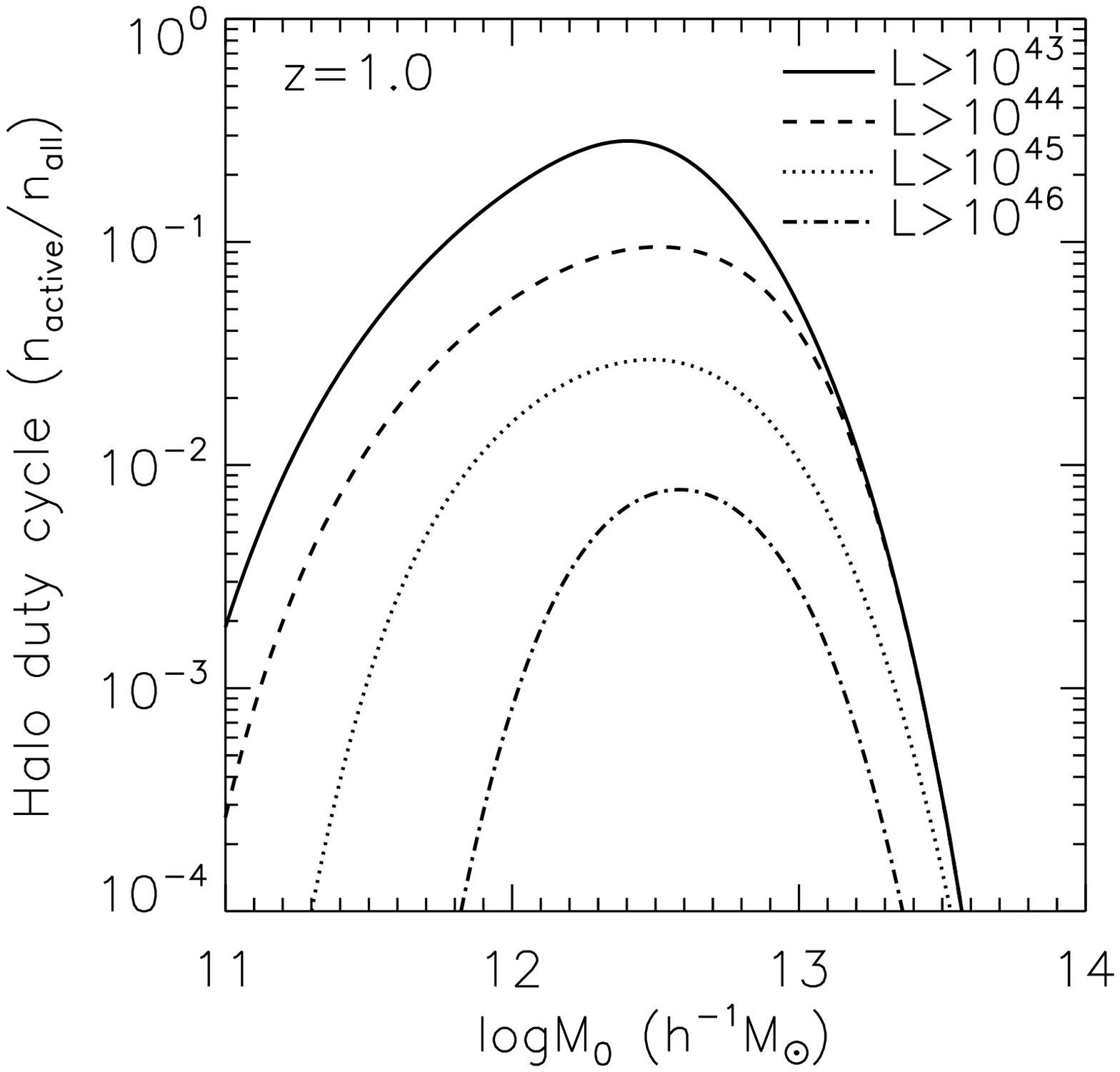}
    \includegraphics[width=0.48\textwidth]{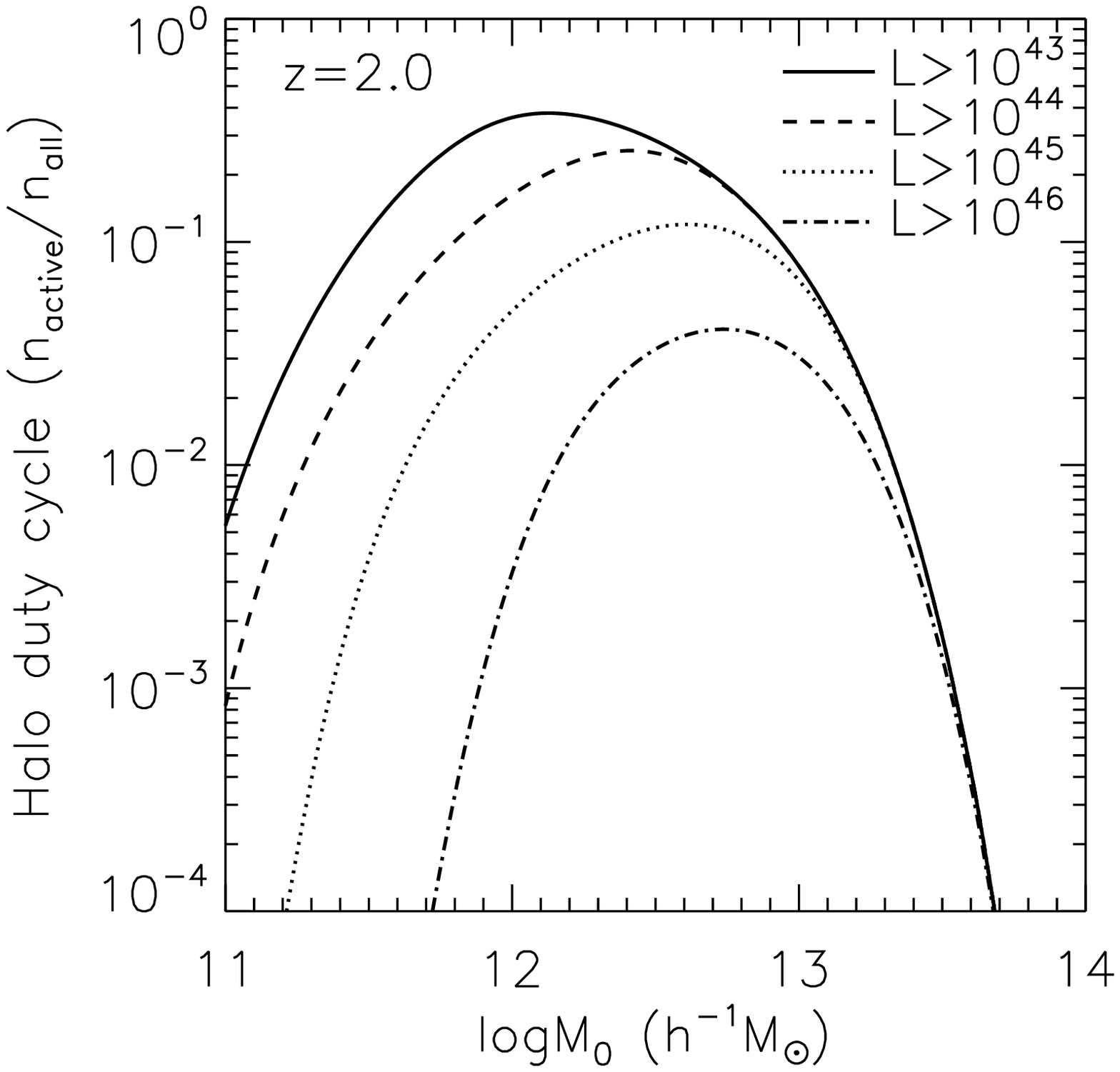}
    \includegraphics[width=0.48\textwidth]{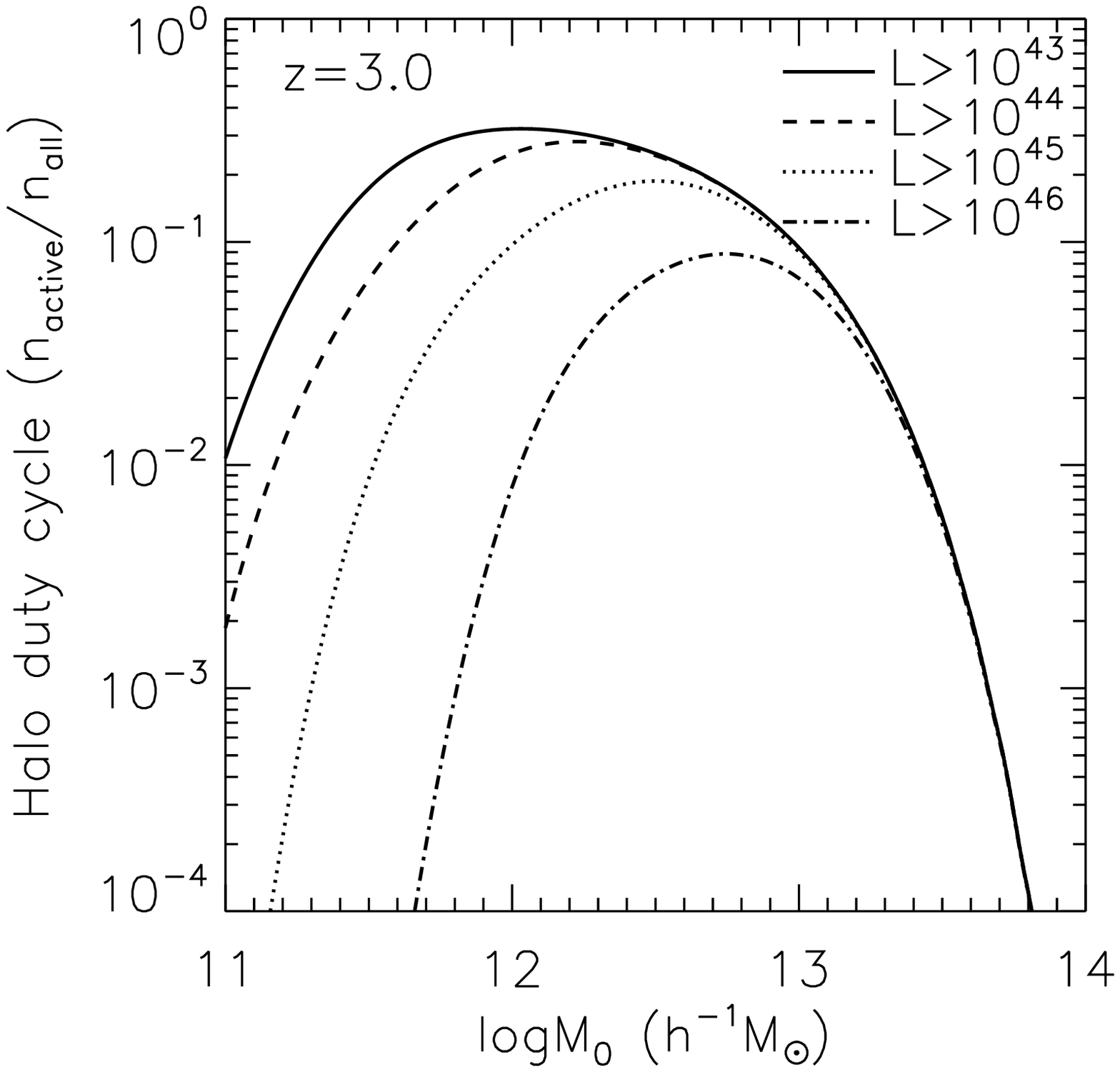}
    \caption{Halo duty cycles, i.e., fraction of halos hosting QSOs brighter than $L_{\rm min}$,
    as function of halo mass, computed using the halo mass function from \citet{Sheth_Mo_Tormen_2001} and the active
    halo mass function from Eqn.\ (\ref{eqn:active_hmf}).}
    \label{fig:act_halo}
\end{figure*}

Fig.\  \ref{fig:act_halo} shows the halo duty cycles, defined as
the ratio of the number density of active halos hosting QSOs
brighter than $L_{\rm min}$ to that of all halos, as function of
halo mass. We have used the \citet{Sheth_Mo_Tormen_2001} halo mass
function and Eqn.\ (\ref{eqn:active_hmf}) for the active halo mass
function. For quasars ($L>10^{45}\,{\rm ergs^{-1}}$) and for
typical halo mass $M_0\sim 2\times 10^{12}\,h^{-1}M_\odot$, the
duty cycle is $\sim 0.15,0.1,0.03,0.01$ at $z=3,2,1,0.5$.

\begin{figure}
  \centering
    \includegraphics[width=0.48\textwidth]{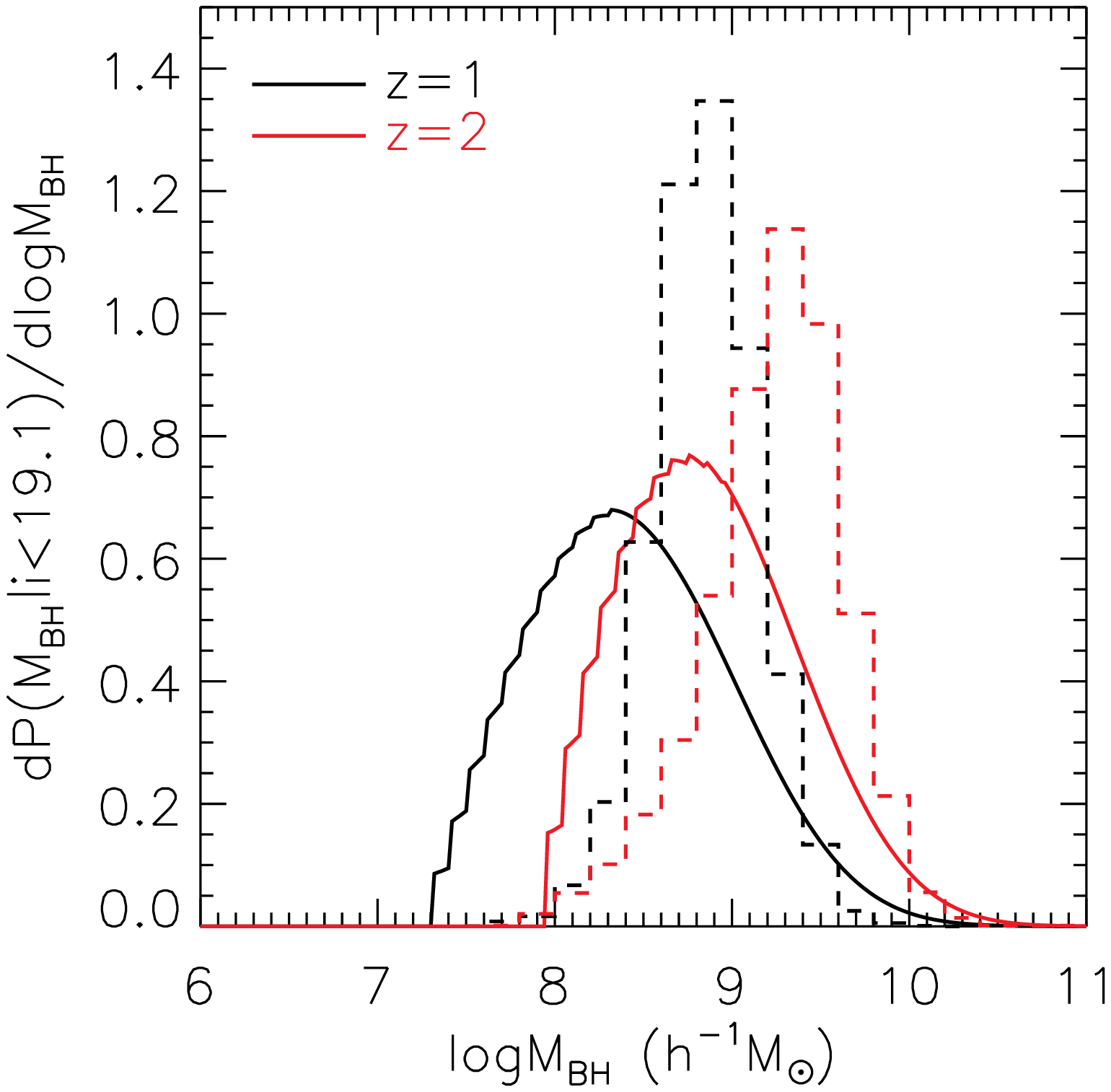}
    \caption{Distributions of BH masses in quasars at two redshifts. The solid lines are the predictions
    of our reference model for
    $i<19.1$ (i.e., the flux limit in the main SDSS quasar catalog) quasars. The dashed lines show the
    distributions of {\em virial} BH masses from \citet{Shen_etal_2008b}. }
    \label{fig:BH_dist}
\end{figure}


We can also compare the model predicted BH mass distributions for
quasars within certain luminosity ranges with the virial BH mass
estimates. For this purpose we use the virial BH mass estimates
from \citet{Shen_etal_2008b} for the SDSS DR5 quasar catalog
\citep{Schneider_etal_2007}. These optical quasars are
flux-limited to $i=19.1$ at $z\lesssim 3$ and we have used eqn.
(1) in \citet{Shen_etal_2009a} to convert $i$-band magnitude to
bolometric luminosity. For quasars/AGNs where the SMBHs are still
actively accreting, what we observe is the instantaneous BH mass
rather than the relic BH mass. In Fig.\ \ref{fig:BH_dist} the
solid lines show the model predictions for the BH mass
distributions of SDSS quasars at $z=1$ and $z=2$, weighted by the
LF; the dashed lines show the distributions based on virial BH
masses. It is remarkable that not only the distributions of our
model predictions are broader, but also the peaks are shifted to
lower masses ($\sim 0.6$ dex) compared with those from virial mass
estimates -- as already discussed extensively in \citet[][i.e.,
the Malmquist-type bias in virial mass
estimates]{Shen_etal_2008b}.


\section{Discussion}\label{sec:disc}

\subsection{Comparison with Previous
Work}\label{subsec:comparison}

The merger basis of our framework, as advocated by many authors
\citep[e.g.,][]{Kauffmann_Haehnelt_2000,Wyithe_Loeb_2002,Wyithe_Loeb_2003,
Volonteri_etal_2003}, provides a physical origin for the QSO
population, and distinguishes the current study from other works
which focus on BH growth using the QSO LF as an input
\citep[e.g.,][]{Yu_Tremaine_2002,
Yu_Lu_2004,Yu_Lu_2008,Marconi_etal_2004,Merloni_2004,Shankar_etal_2004,Shankar_etal_2009a}.
Our simple framework, although semi-analytical in nature,
accommodates a wide range of updated and new observations of QSO
statistics; these new observations include quasar clustering and
Eddington ratio distributions. Most of the early quasar models
\citep[e.g.,][]{Haiman_Loeb_1998,Kauffmann_Haehnelt_2000,Wyithe_Loeb_2002,Wyithe_Loeb_2003,
Volonteri_etal_2003} focused mainly on the luminosity function
(partly because other observations were not available at that
time), and most of them assumed simplified light curve models
which were unable to reproduce the observed Eddington ratio
distributions.

\citet{Hopkins_etal_2008a} presented a merger-based quasar model
that utilizes a variety of quasar observations for comparison,
including the latest clustering measurements. Our model framework
is different from theirs in two major aspects: 1) they estimated
the major-merger rate from the combination of empirical halo
occupation models of galaxies and merging timescale analysis,
while we used directly the halo merger rate from simulations --
both approaches have their own advantages and disadvantages; 2)
the light curve in their model is extracted from their
merger-event simulations \citep{Hopkins_etal_2005a}, while we have
adopted a parametrization of the light curve which is fit by
observations. Their light curve model is clearly more
physically-motivated than ours, yet it needs to be confirmed in
future simulations with higher resolution and better
understandings of BH accretion physics. On the other hand, the
virtue of our framework is that it allows fast and easy
estimations of model predictions and parameter adjustments to fit
updated observations.

\subsection{Caveats in the Reference Model}\label{subsec:disc1}
As already mentioned in \S\ref{sec:ref_model}, our fiducial model
is not an actual $\chi^2$ fit to the overall observations (LF,
clustering, and Eddington ratio distributions) because of the
ambiguity of assigning relative weights to individual
observational data sets. Instead, we have experimented with
varying the model parameters within reasonable ranges, to achieve
a global ``good'' (as judged by eye) fit to observations. If
consider only the LF as observational constraints, there are model
degeneracies between the luminosity decaying rate $\alpha$ and the
normalization of the mean $L_{\rm peak}-M_0$ relation $C$, i.e.,
if QSO luminosity decays more slowly, the typical host halos need
to shift to more massive (and less abundant) halos in order not to
overpredict the LF; likewise, if the scatter around the mean
$L_{\rm peak}-M_0$ relation $\sigma_L$ increases, we can reduce
$C$ to match the LF. However, these degeneracies are broken once
the clustering observations are taken into account. A large
scatter $\sigma_L$ or slow varying light curve (small values of
$\alpha$) cannot fit the large bias at high redshift and the
luminosity dependence of clustering. Our model is also more
complicated than previous models
\citep[e.g.,][]{Haiman_Loeb_1998,Kauffmann_Haehnelt_2000,Wyithe_Loeb_2002,
Wyithe_Loeb_2003,Volonteri_etal_2003,Shankar_etal_2009a} in the
sense that we have more parameters, which is required for a
flexible enough framework to accommodate a variety of
observations.

For our fiducial model we have used the halo merger rate as the
proxy for the QSO-triggering rate. Alternatively, if we use the
subhalo merger rate (the delayed version of halo merger rate;
\S\ref{subsec:intro_gal_merger}), it makes little difference below
$z=3$, but it further underestimates the LF at $z>3$, as expected
from Fig.\ \ref{fig:tau_merger}. The bolometric LF data we used
here has uncertainties both from measurements and bolometric
corrections at the $\sim 20-30\%$ level
\citep[cf.,][]{Hopkins_etal_2007a,Shankar_etal_2009a}, not enough
to reconcile the discrepancy at $z>4.5$. There are several ways to
modify our model to match observations of LF at $z>4.5$: a)
decrease the threshold $\xi_{\rm min}$ above which mergers can
trigger QSO activity; b) modify the $L_{\rm peak}-M_0$ relation
such that at fixed peak luminosity, QSOs shift to even smaller
halos at $z>4.5$, which can be achieved by including some higher
order terms in the redshift evolution of $C$; and c) increase the
scatter $\sigma_L$ so that more abundant low mass halos can
contribute to the LF by up-scattering. As an example of such a
model, we modify the redshift evolution of the mean $L_{\rm
peak}-M_0$ relation at high redshift such that $C\rightarrow C +
\log[(\frac{1+z}{4.5})^{7/2}]$ for $z>3.5$. With this additional
term of evolution in the mean $L_{\rm peak}-M_0$ relation, the
typical host halo shifts to lower masses ($M_0\sim 4\times
10^{11}\, h^{-1}M_\odot$ for $\langle\log L/{\rm
ergs^{-1}}\rangle=46$ at $z=6$, compared to $M_0\sim
10^{12}\,h^{-1}M_\odot$ in our fiducial model), and the resulting
LF (dashed lines in Fig.\  \ref{fig:LFs}) fit the observations
well. This modification has little effects on the LF at $z<3.5$,
as well as other predicted QSO properties. An alternative
parametrization of the redshift evolution in $C$ is further
discussed in \S\ref{subsec:disc2}.

However, there are no independent constraints on the mass of halos
hosting quasars at $z>4.5$ (such as those inferred from quasar
clustering), and the merger scenario of quasar activity may be
different at such high redshift. Hence we do not attempt to fully
resolve this issue in this paper.

There is also slight tension between quasar clustering and LF at
$z\sim 4$ in our model, as already noted by several studies
\citep[][]{White_etal_2008,Wyithe_Loeb_2008,Shankar_etal_2009b}.
On the one hand, we need smaller halo masses to account for quasar
abundance; we also need larger halos to account for the strong
clustering on the other. To fully resolve this issue we need
better understanding of the halo bias, {\em and reliable fitting
formulae for it}, derived from simulations for the relevant mass
and redshift ranges\footnote{For instance, it has been suggested
that in addition to mass, halo clustering also depends on
concentration, assembly history and recent merger activity
\citep[the so-called assembly bias, e.g., ][]{Gao_etal_2005,
Wechsler_etal_2006,Wetzel_etal_2007}. If recently merged halos (as
quasar hosts) have larger bias than average for the same halo
mass, then it may reconcile the slight tension between reproducing
both the LF and quasar clustering at high redshift
\citep[e.g.,][]{Wyithe_Loeb_2008}, although current estimate of
this enhancement is only on the level of $\sim 10\%$
\citep[e.g.,][]{Wetzel_etal_2007}.}, as well as better
measurements of quasar clustering at high redshift with future
larger samples. Nevertheless, our fiducial model is still
consistent with both LF and clustering observations within the
errors.

Our model also underpredicts the LF at the low luminosity end
($L<10^{45}\,{\rm ergs^{-1}}$) at $z<0.5$. This is somewhat
expected, since our model does not include contributions from AGNs
triggered by mechanisms other than a major merger. The fuel budget
needed to feed a low luminosity AGN ($L<10^{45}\,{\rm ergs^{-1}}$)
is much less stringent than that for bright quasars. Therefore
secular processes (i.e., gas inflows driven by bars, tidal
encounters, stochastic accretion, as well as minor mergers), while
not as violent and efficient as major mergers, provide viable
means to fuel AGNs at low activity levels. At $z<0.5$, there are
evidence that some low luminosity AGN ($L\sim 10^{43-44}\,{\rm
ergs^{-1}}$, powered by intermediate-mass BHs $M_\bullet\sim
10^6\,M_\odot$) hosts have no classical bulges\footnote{Classical
bulges (including ellipticals and bulges in early-type disk
galaxies) are presumably formed via mergers and they follow the
fundamental plane of elliptical galaxies
\citep[e.g.,][]{Bender_etal_1992}. On the other hand, secular
processes can build up the so-called {\em pseudobulges}, which
have distinct structural properties from classical bulges
\citep[][]{Kormendy_Kennicutt_2004}. While pseudobulges can host
central BHs, there are some evidence that the bulge-BH mass
relations for pseudobulge systems are offset from that for
classical bulge systems \citep[e.g.,][]{Hu_2008,Greene_etal_2008},
indicating that secular processes are less efficient in building
BHs.}, which indicates that secular processes are responsible for
triggering these low-luminosity AGN activity and BH growth
\citep[e.g.,][and references therein]{Greene_etal_2008}. There are
also observational implications that the bulk of BH growth has
shifted from the most massive BHs ($M_{\bullet}>10^8\,M_\odot$) at
high redshift ($z\lesssim 2$) to low mass BHs
($M_\bullet<10^8\,M_\odot$) locally
\citep[e.g.,][]{McLure_Dunlop_2004,Heckman_etal_2004}, along with
the cosmic downsizing in luminosity function evolution
\citep[e.g.,][]{Steffen_etal_2003,Ueda_etal_2003,Hasinger_etal_2005,Hopkins_etal_2007a,Bongiorno_etal_2007}.
The typical transition from merger-driven BH growth to
secularly-driven BH growth likely occurs around BH mass $\sim
10^7\,M_\odot$ \citep[e.g.,][]{Hopkins_etal_2008b}, corresponding
to $L\sim 10^{45}\,{\rm ergs^{-1}}$ at the Eddington limit. Since
the specific merger rate increases towards higher redshift, while
the rate of secular processes is almost constant with time and
these processes are relatively slow and inefficient for BH growth,
it is conceivable that secular processes will only become
important at late times ($z<0.5$ for instance) in building up the
low mass end of the SMBH population. The implementation of
secularly-driven AGN activity in our model will be presented in
future work.

Finally, we mention that our model predicts a turnover in the LF
at $L\lesssim 10^{44}\,{\rm ergs^{-1}}$ at $z\gtrsim 3$. This
simply reflects the lower mass cutoff at $M_{\rm min}=3\times
10^{11}\,h^{-1}M_\odot$ in our model. Future deeper surveys for
low luminosity AGNs at $z\gtrsim 3$ are necessary to probe this
luminosity regime, and to impose constraints on how efficiently
SMBHs can form in low mass halos.

\subsection{The Effects of BH
Coalescence on the BHMF}\label{subsec:BH_Coalescence}

The BHMFs discussed in \S\ref{sec:ref_model} and Fig.\
\ref{fig:act_BMF} are the mass functions from accretion only,
i.e., we have neglected the effects of BH coalescence. As
discussed in \S\ref{subsec:SMBH_demo}, BH coalescence will
redistribute the BH mass function but does not change the total BH
mass density (neglecting mass loss via gravitational radiation)
since essentially all the mass ended up in BHs were accreted.
There are two routes in which the coalescence with pre-existing
BHs might become important within our model framework\footnote{In
contrary to the two cases discussed below, we assume that a major
merger between two spirals will lead to negligible BH mass
contribution from the pre-existing BHs, since in our model
setting, spiral galaxy has not yet experienced a major merger and
hence significant BH growth.}. First, both galaxies in the merging
pair of halos are elliptical, i.e., they already experienced a QSO
phase in the past and formed massive nuclear BHs. In this case the
current merger event will be a dry merger and form a SMBH binary
without triggering a new QSO. Second, during the major merger, one
galaxy is elliptical and the other one is spiral, in which case a
QSO will be triggered and the mass of the old BH of the previous
elliptical will add to the new BH system.

A complete exploration of these routes and their consequences
(including the effects of BH ejection, mass loss through
gravitational radiation, etc.) can be better achieved with Monte
Carlo realizations of halo merger trees and our model
prescriptions for QSO triggering and BH growth, which we plan to
investigate in a future paper. Here we can approximately estimate
the maximal impact of BH coalescence on the BHMFs in the two
cases, assuming that BH coalescence always occurs within a Hubble
time and all the mass in the pre-existing BHs is added to the
final BH, i.e., no BH ejection or mass loss via gravitational
radiation. We focus on the high-mass end ($M_\bullet\gtrsim 10^8\,
M_\odot$) of the local BHMF since our model prediction is
incomplete at the lower-mass end.

Although the {\em dry merger} case is not subject to the major
merger condition for QSO-triggering, we still restrict to $\xi\ge
\xi_{\rm min}$ here because in minor mergers: 1) it will take too
long for the two pre-existing BHs to become a close binary, and 2)
the mass increment due to BH coalescence is insignificant.
Observational determination of the major dry merger rate is
difficult, and the current best estimate is: on average,
present-day spheroidal galaxies with $M_V<-20.5$ (corresponding to
$M_{\bullet}\gtrsim 10^8\,M_\odot$) have undergone $0.5-2$ major
dyr mergers since $z\sim 0.7$ \citep[][]{Bell_etal_2006}. If we
assume all the $M_{\bullet}\gtrsim 10^8\,M_\odot$ BHs undergo one
$1:1$ dry merger after the QSO phase, the local BHMF will
redistribute as the yellow line in Fig.\  \ref{fig:act_BMF}. In
this case the abundance of the most massive ($M_\bullet>$ a few
$\times 10^9\,M_\odot$) BHs is enhanced by up to a factor of $\sim
2$ at $M_\bullet=10^{10}\,h^{-1}M_\odot$.

In the {\em half-dry major merger} case, the fraction of the
pre-existing BH mass to the final BH mass is
$(1+\xi^\prime)^{-5/3}\sim 0.07-0.7$, where $1/4\le
\xi^\prime\le4$ is the major merger mass ratio (in our reference
model) between the two halos. This is because the final BH mass
after a QSO phase scales as the $5/3$ power to the halo mass in
our model. Averaging over possible values of $\xi^\prime$, the
fractional increment due to the pre-existing BH is $\sim 30\%$. If
all QSO-triggering mergers are this kind of {\em half-dry} event,
the predicted $z=0$ BHMF (red line in Fig.\ \ref{fig:act_BMF})
will redistribute from lower mass to higher mass (the green line).
The enhancement at the high-mass end of the local BHMF is
comparable to the dry major merger case. In practice
QSO-triggering mergers can occur between two spirals, hence the
actual correction due to pre-existing BHs in {\em half-dry}
mergers should be smaller.

Combining these two cases we conclude that the impact of BH
coalescence on the BHMF is probably insignificant compared with
other uncertainties and systematics in current observations and
our model framework. Similar conclusions were also achieved in
several independent work
\citep[][]{Volonteri_etal_2003,Yu_Lu_2008,Shankar_etal_2009a}
albeit with difference in details.

\subsection{Implications for BH Scaling
Relations}\label{subsec:disc2}

In our formalism there is a generic scaling relation between the
relic BH mass and halo mass, which has the same slope and scatter
as the $L_{\rm peak}-M_0$ relation (from Eqns. \ref{eqn:Lpeak_M0}
and \ref{eqn:relic_mass1}):
\begin{equation}\label{eqn:Mbh_M0}
\frac{M_{\rm \bullet,relic}}{10^8\, h^{-1}M_\odot}\approx
0.6(1+z)^{\beta_1}\bigg(\frac{M_0}{10^{12}\,h^{-1}M_\odot}\bigg)^{5/3}\
.
\end{equation}
The local $M_{\bullet}-M_0$ relation for dormant BHs reported in
\citet[][]{Ferrarese_2002} and \citet{Baes_etal_2003} has a slope
in the range $\sim 1.3-1.8$, and a normalization lower by a factor
of $\sim 3-5$ than our predictions. This is probably due to the
fact that we neglected continued growth of halos by minor mergers
and diffuse matter accretion since the major merger event, if most
of the local massive BHs were assembled at $z\ga 1$ (as our model
predicts). The fact that we did not impose a cutoff in the LC
(\ref{eqn:LC_model}) may also lead to overly massive BHs. We note
that although the normalization (and perhaps slope as well) of the
local $M_{\bullet}-M_0$ relation may depend on galaxy
morphological type
\citep[e.g.,][]{Zasov_etal_2004,Courteau_etal_2007,Ho_2007},
early-type galaxies (S0 and ellipticals) appear to occupy the
upper envelope in the $M_{\bullet}-M_0$ relation. Our model
scaling relations are only valid for early type galaxies which are
presumably merger remnants.


However, it should be pointed out that the BH-halo scaling
relation relies on the assumed $L_{\rm peak}-M_0$ relation. The
simple prescription for its evolution in our reference model
already shows some difficulties in reproducing the QSO LF at $z\ga
4.5$ (Fig.\ \ref{fig:LFs}; see \S\ref{subsec:disc1}). Hence our
fiducial model is not very appropriate for predicting the redshift
evolution of the BH-halo scaling relation. To make this point more
clear, let us consider a more physically-motivated prescription
for the $L_{\rm peak}-M_0$ relation in which $L\propto V_{\rm
vir}^5$, where $V_{\rm vir}$ is the halo virial velocity (Eqn.\
\ref{eqn:Vvir}), and the normalization of the $L_{\rm peak}-M_0$
relation evolves as \citep[][]{Wyithe_Loeb_2002,Wyithe_Loeb_2003}:
\begin{equation}\label{eqn:WL03}
C(z)=C(z=0)+\displaystyle\frac{5}{2}\log(1+z)+\frac{5}{6}\log\bigg[\frac{\Delta_{\rm
vir}(z)}{\Delta_{\rm vir}(0)}\bigg]\ ,
\end{equation}
e.g., the evolution in $C(z)$ is more rapid than our fiducial
setting. With this new implementation for the $L_{\rm peak}-M_0$
relation, we found that a good global fit can be achieved with the
following parameter adjustments (other parameters are the same as
in our reference model): $M_{\rm quench}=3\times 10^{12}\,
h^{-1}M_\odot$, $C(z=0)=\log(0.8\times 10^{45})-12\gamma$ and
$\sigma_{L}=0.4(1+z)^{-1/2}$. Since the redshift evolution in the
normalization $C(z)$ is now faster, the starting value $C(0)$ is
reduced; consequently the exponential upper cut of halo mass,
$M_{\rm quench}$, increases in order to account for QSO counts at
low redshift. The scatter in the $L_{\rm peak}-M_0$ relation needs
a redshift evolution to achieve adequate fits for both clustering
and LF over a wide redshift range. The predicted LF is shown as
dotted lines in Fig.\ \ref{fig:LFs}, which does a much better job
at $z\ga 4.5$ than our fiducial model. Other predicted properties
are slightly degraded (but still are reasonably good fits to
observations) than those predicated by our fiducial model.

This new $L_{\rm peak}-M_0$ relation, together with the
approximation that the halo virial velocity $V_{\rm vir}\approx
v_c$, the galaxy circular velocity, and the assumption that the
local $v_c-\sigma$ relation \citep[][]{Ferrarese_2002} does not
evolve, result in a constant $M_\bullet-\sigma$ relation
\citep[][]{Wyithe_Loeb_2003}. Neglecting the scatter, the local
mean $v_c-\sigma$ relation in \citet{Ferrarese_2002} is:
\begin{equation}\label{eqn:vc_sigma}
\log \bigg(\frac{v_c}{\rm km\,
s^{-1}}\bigg)=0.84\log\bigg(\frac{\sigma}{\rm km\, s^{-1}}\bigg) +
0.55\ .
\end{equation}
Thus the new $L_{\rm peak}-M_0$ relation predicts a constant
$M_{\bullet}-\sigma$ relation (e.g., Eqns.\ \ref{eqn:Lpeak_M0},
\ref{eqn:BH_growth_M}, \ref{eqn:relic_mass1}, \ref{eqn:WL03}, and
\ref{eqn:Mvir_Vvir}):
\begin{equation}\label{eqn:M-sigma}
\frac{M_\bullet}{10^8\,M_\odot}=
0.9\sim5.0\times\bigg(\frac{\sigma}{200\, {\rm km\,
s^{-1}}}\bigg)^{4.2}\ ,
\end{equation}
with a slope and normalization (bounded by $M_{\bullet,\rm peak}$
and $M_{\bullet,{\rm relic}}$) consistent with local estimates
\citep[][]{Gebhardt_etal_2000a,Ferrarese_Merritt_2000,Tremaine_etal_2002,Lauer_etal_2007a}.
This consistency, however, is built on a couple of assumptions and
approximations, and neglecting successive evolution after the
self-regulation of BH and bulge growth. Given all these
complications, it is beyond the scope of the current study to
fully settle this issue. We simply remind the reader that a
non-evolving $M_\bullet-\sigma$ relation is generally allowed
within our framework.


\section{Conclusions}\label{sec:con}

We have developed a general cosmological framework for the growth
and cosmic evolution of SMBHs in the hierarchical merging
scenario. Assuming that QSO activity is triggered by major mergers
of host halos, and that the resulting light curve follows a
universal form with its peak luminosity correlated with the
(post)merger halo mass, we model the QSO LF and SMBH growth
self-consistently across cosmic time. We tested our model against
a variety of observations of SMBH statistics: the QSO luminosity
function, quasar clustering, quasar/AGN BH mass and Eddington
ratio distributions. A global good fit is achieved with reasonable
parameters. We summarize our model specifics as follows:
\begin{itemize}

\item The QSO-triggering rate is determined by the $\xi\ge 0.25$
halo merger rate with exponential cutoffs at both the low and high
halo mass ends $M_{\rm min}=3\times 10^{11}\,h^{-1}M_\odot$,
$M_{\rm max}(z)=10^{12}(1+z)^{3/2}\,h^{-1}M_\odot$.

\item The universal light curve follows an initial exponential
Salpeter growth with constant Eddington ratio $\lambda_0=3$ for a
few $e$-folding times to reach the peak luminosity $L_{\rm peak}$,
which is correlated with the (post)merger halo mass as $\langle
L_{\rm peak}\rangle=6\times
10^{45}(1+z)^{1/3}(M_0/10^{12}\,h^{-1}M_\odot)^{5/3}\,{\rm
ergs^{-1}}$, with a log-normal scatter $\sigma_L=0.28$ dex. It
then decays as a power-law with slope $\alpha=2.5$.

\end{itemize}

Our simple model successfully reproduces the LF, quasar
clustering, and Eddington ratio distributions of quasars and AGNs
at $0.5<z<4.5$, supporting the hypothesis that QSO activity is
linked to major merger events within this redshift range. However,
there are still many unsettled issues. Below we outline several
possible improvements of our simple model, which will be addressed
in future work.

Our model under-predicts the LF at the faint luminosity end
$L<10^{45}\,{\rm ergs^{-1}}$ at $z<0.5$, which is linked to the
growth of the less massive $\lesssim 10^7\,M_\odot$ SMBHs. This is
indicative of a population of low luminosity AGNs triggered by
mechanisms other than major mergers at the low redshift universe
-- either by minor mergers or secular processes. We need to
incorporate this ingredient in our SMBH model, in order to match
the local BHMF at the low mass end ($M_{\bullet}\lesssim
10^7\,M_\odot$).

In our modeling we have neglected the possibilities of a
closely-following second major merger event and the triggering of
two simultaneous QSOs during a single major merger event.
Therefore our model does not include more than one QSOs within a
single halo. We will use Monte-Carlo realizations of halo merger
trees to assess the probability of such rare occurrences and see
if they can account for the small ($\lesssim 0.1\%$)
binary/multiple quasar fraction observed
\citep[][]{Hennawi_etal_2006,Hennawi_etal_2009,Myers_etal_2008}.

Our model can be improved to include the radio loudness of QSOs as
well. If radio loudness requires both a massive host halo (to
provide the hot IGM) and a massive SMBH (to launch the kinetic
jet), we can statistically populate radio-loud QSOs in halos
within our model framework. It can be tested against the
clustering of radio-loud quasars \citep[e.g.,][]{Shen_etal_2009a}
and the radio-loud fraction as function of luminosity and redshift
\citep[e.g.,][and references therein]{Jiang_etal_2007a}.

\acknowledgements We thank the anonymous referee, Francesco
Shankar, Michael Strauss, Scott Tremaine, and Martin White for
constructive comments that have greatly improved the manuscript.
We are grateful to Francesco Shankar for pointing out an error in
the merger rate equation (4) in an earlier version of the
manuscript. This work was supported by NSF grant AST-0707266.


\appendix

\section{Dark Matter Halos}
\label{app:B}

All dark matter halos are assumed to have a spherical NFW profile
\citep[][]{NFW_1997}:
\begin{equation}
\rho_{\rm NFW}(r)=\frac{\rho_s}{(r/r_s)(1+r/r_s)^2}\ ,
\end{equation}
where $r_s$ and $\rho_s$ are the characteristic scale and the
density at this scale.

The virial mass and virial radius are related by
\citep[][]{Bullock_etal_2001a}:
\begin{equation}
M_{\rm vir}\equiv \frac{4\pi}{3}\Delta_{\rm vir}\rho_ur_{\rm
vir}^3\ ,
\end{equation}
where $\Delta_{\rm vir}(z)\approx (18\pi^2+82x-39x^2)/\Omega(z)$
with $x\equiv \Omega(z)-1$ \citep[][]{Bryan_Norman_1998} is the
spherical overdensity relative to the background {\em matter}
density $\rho_u$ and $\Omega(z)=\displaystyle
[1+\Omega_\Lambda(1+z)^{-3}/\Omega_0]^{-1}$. Note there is the
slight difference in defining the virial radius in
\citet{Bullock_etal_2001a} and \citet{NFW_1997} where the latter
uses $r_{200}$ (the radius corresponding to a spherical
overdensity 200 times the {\em critical} density) to define the
virial radius. Both definitions of virial radius are frequently
used in studies on the galaxy merger time scale within merged dark
matter halos: \citet[][]{Jiang_etal_2008},
\citet{Stewart_etal_2008} used the former definition, while
\citet[][]{Boylan-Kolchin_etal_2008}, \citet{Wetzel_etal_2008a}
used the latter.

The enclosed mass within an NFW profile truncated at $r_o$ is
\begin{equation}
M(r_o)=\int_0^{r_o}dr 4\pi r^{2}\rho_{\rm
NFW}(r)=4\pi\rho_sr_s^3\bigg[\ln(1+c)-\frac{c}{1+c}\bigg]\ ,
\end{equation}
where $c\equiv r_o/r_s$. Therefore we have
\begin{equation}
M_{\rm vir}=M(r_{\rm vir})=4\pi\rho_sr_s^3\bigg[\ln(1+c_{\rm
vir})-\frac{c_{\rm vir}}{1+c_{\rm vir}}\bigg]
\end{equation}
where $c_{\rm vir}\equiv r_{\rm vir}/r_s$ is the usual definition
of the concentration parameter. The mean relation between $M_{\rm
vir}$ and $c_{\rm vir}$ is given in \citet{Bullock_etal_2001a}.

The virial velocity (usually defined as the circular velocity at
the virial radius) $V_{\rm vir}$, and the maximum circular
velocity at $r_{\rm max}\approx 2.16r_s$ are
\citep[][]{Bullock_etal_2001a}:
\begin{equation}\label{eqn:Vvir}
V_{\rm vir}^2\equiv V_{c}^2(r_{\rm vir})= \frac{GM_{\rm
vir}}{r_{\rm vir}}\ ,\qquad \frac{V_{\rm max}^2}{V_{\rm
vir}^2}\approx \frac{0.216c_{\rm vir}}{\ln(1+c_{\rm vir})-c_{\rm
vir}/(1+c_{\rm vir})}\ .
\end{equation}
This implies that the relation between virial mass $M_{\rm vir}$
and virial velocity $V_{\rm vir}$ is:
\begin{equation}\label{eqn:Mvir_Vvir}
M_{\rm vir}(z)=\bigg[\frac{4\pi}{3}\Delta_{\rm
vir}(z)\rho_0\bigg]^{-1/2}(1+z)^{-3/2}G^{-3/2}V_{\rm
vir}^3=1.37\times
10^{12}\Omega_0^{-1/2}h^{-1}M_\odot\bigg(\frac{V_{\rm
vir}}{200\,{\rm km\,
s^{-1}}}\bigg)^3(1+z)^{-3/2}\bigg[\frac{\Delta_{\rm
vir}(z)}{\Delta_{\rm vir}(0)}\bigg]^{-1/2}\ ,
\end{equation}
where $\rho_0=2.78\times 10^{11}\Omega_0h^2M_\odot{\rm Mpc}^{-3}$
is the $z=0$ mean matter density.

The DM halo dynamical time $\tau_{\rm dyn}$ is usually defined as
$r_{\rm vir}/V_{\rm vir}$:
\begin{equation}
\tau_{\rm dyn}\equiv \frac{r_{\rm vir}}{V_{\rm vir}}=1.4\times
10^{10}\,{\rm yr}\times [\Omega_0h^2\Delta_{\rm
vir}(1+z)^3]^{-1/2}\ .
\end{equation}

For simplicity we have neglected the difference between $M_{\rm
vir}$ and the {\em friends-of-friends} mass $M_{\rm fof}$ (with a
link length $b=0.2$) throughout the paper. But we give an
approximate conversion formula below for
completeness: 
\begin{equation}
\frac{M_{\rm fof}}{M_{\rm vir}}=\frac{\ln(1+c_{\rm fof})-c_{\rm
fof}/(1+c_{\rm fof})}{\ln(1+c_{\rm vir})-c_{\rm vir}/(1+c_{\rm
vir})}\ ,
\end{equation}
where $c_{\rm fof}=r_{\rm fof}/r_s$ can be solved via the
following equation:
\begin{equation}
c_{\rm fof}(1+c_{\rm fof})^2=\frac{2\pi b^3\Delta_{\rm
vir}}{9}\frac{c_{\rm vir}^3}{\ln(1+c_{\rm vir})-c_{\rm
vir}/(1+c_{\rm vir})}\ ,
\end{equation}
where we have used the fact that the density at $r_{\rm fof}$ is
$\rho_{\rm NFW}(r_{\rm fof})\approx 3\rho_u/(2\pi b^3)$.

\clearpage
\bibliography{D:/Research/bib/smbhrefs}

\end{document}